\documentclass[useAMS,usenatbib]{mn2e}

\usepackage{graphicx,amsmath,eufrak,rotating}
\usepackage{natbib}
\usepackage{fixltx2e}
\usepackage{hyperref}
\usepackage{rotating}
\usepackage{lscape}

\newcommand{\msun}{M$_{\sun}$}

\newcommand\tiri{{\sc TiRiFiC}}
\newcommand\HI{H{\sc i}}

\newcommand\rc{{\sc rotcur}}
\newcommand\df{{\sc DiskFit}}
\newcommand\sofia{{\sc SoFiA}}

\newcommand\FAT{{\sc FAT}}

\title[Automated Kinematic Modelling  in Large \HI\ Surveys]{Automated Kinematic Modelling of Warped Galaxy Discs in Large \HI\ Surveys: 3D Tilted Ring Fitting of HI Emission Cubes.}
\author[P. Kamphuis et al.]{P. Kamphuis$^{1}$\thanks{E-mail: peterkamphuisastronomy@gmail.com},  G. I. G. J\'ozsa$^{2,3,4}$, S-.H. Oh$^{5,6}$, K. Spekkens$^7$, N. Urbancic$^7$, \newauthor  P. Serra$^{1}$, B. S. Koribalski$^{1}$, R.-J. Dettmar$^{8}$. \\
$^{1}$CSIRO Astronomy \& Space Science, Australia Telescope National Facility, P.O. Box 76, Epping, NSW 1710, Australia\\
$^{2}$SKA South Africa, Radio Astronomy Research Group, 3rd Floor, The Park, Park Road, Pinelands, 7405, South Africa\\
$^{3}$Rhodes University, Department of Physics and Electronics, Rhodes Centre for Radio Astronomy Techniques \& Technologies, \\
\hspace*{12.5 cm}PO Box 94, Grahamstown, 6140, South Africa\\
$^{4}$Argelander-Institut f\"ur Astronomie, Universit\"at Bonn, Auf dem H\"ugel 71, 53121 Bonn, Germany\\
$^{5}$International Centre for Radio Astronomy Research (ICRAR), Univ. of Western Australia, 35 Stirling Highway, Perth, WA 6009, Australia\\
$^{6}$ARC Centre of Excellence for All-sky Astrophysics (CAASTRO), 44-70 Rosehill Street, Redfern NSW 2016, Sydney, Australia\\
$^{7}$Department of Physics, Royal Military College of Canada, P.O. Box 17000, Station Forces, Kingston, Ontario K7K 7B4, Canada\\
$^{8}$Astronomisches Institut Ruhr-Universit\"at Bochum, Universit\"atstrasse 150, D-44801 Bochum, Germany\\
}
\begin{document}

\date{}

\pagerange{\pageref{firstpage}--\pageref{lastpage}} \pubyear{2013}
\maketitle
\label{firstpage}
\begin{abstract}
Kinematical parameterisations of disc galaxies, employing emission line observations, are indispensable tools for studying the formation and evolution of galaxies. Future large-scale \HI\ surveys will resolve the discs of many thousands of  galaxies, allowing a statistical analysis of their disc and halo kinematics, mass distribution and dark matter content. \\
\indent Here we present an automated procedure which fits tilted-ring models to \HI\  data cubes of individual, well-resolved galaxies. The method builds on the 3D {\it Ti}lted {\it Ri}ng {\it Fi}tting {\it C}ode (\tiri) and is called \FAT\ (Fully Automated \tiri).\\
\indent  To assess the accuracy of the code we apply it to a set of 52 artificial galaxies and 25 real galaxies from the Local Volume \HI\ Survey (LVHIS). Using LVHIS data, we compare our 3D modelling to the 2D modelling methods \df\ and \rc.\\
\indent A conservative result is that FAT accurately models the kinematics and the morphologies of galaxies with an extent of eight beams across the major axis in the inclination range 20\degr-90\degr\ without the need for priors such as disc inclination. When comparing to 2D methods we find that  velocity fields cannot be used to determine  inclinations in galaxies that are marginally resolved.  We conclude that with the current code tilted-ring models can be produced in a fully automated fashion. This will be essential for future  \HI\ surveys, with the Square Kilometre Array and its pathfinders,  which will allow us to model the gas kinematics of many thousands of well-resolved galaxies. Performance studies of \FAT\ close to our conservative limits, as well as the introduction of more parameterised models will open up the possibility to study even less resolved galaxies.\\

\end{abstract}

\begin{keywords}
galaxies: ISM, galaxies: kinematics and dynamics, galaxies: structure, methods: data analysis, surveys
\end{keywords}

\section{Introduction}
The accurate description of the kinematics of galaxies is crucial to get insight into how they form and evolve during their lifetimes. The Doppler-shifted \HI\ line is one of the main tracers of these kinematics; \HI\ is ubiquitous, largely unaffected by absorption and often extends far further out than other probes. As such, parameterisation of the \HI\ distribution and dynamics has been done for several decades, mostly on an individual galaxy basis and in a highly interactive fashion.\\
\indent Successful and objective kinematic parametrisation can lead to a wealth of understanding in galaxy formation and evolution. For example,  it can be used to investigate the origin of warps \citep[e.g.][]{Jozsa2007b, vdKruit2011, Haan2014} and with that, gas accretion via intrinsic and large-scale spin alignment \citep{Popping2014}, but also the study of dark and baryonic matter distributions \citep{Swaters2012, Swaters2014}.\\
\indent Current \HI\ surveys such as  the Westerbork \HI\ Survey of Irregular and Spiral galaxies (WHISP) \citep{vdHulst2002}, the Local Volume \HI\ Survey (LVHIS) \citep{Koribalski2010}, DiskMass \citep{Martinsson2011}, ATLAS$^{{\rm 3D}}$ \citep{Serra2012} and the Blue disc Project \citep{Wang2013} have already observed tens to hundreds of galaxies for which kinematic parametrisation can be obtained. Although galaxy models can still be produced interactively for this number of systems \citep{Swaters2012}, the need for automation is obvious. \\
\indent This need  is further emphasised when looking at the planned \HI\ surveys for the Australian SKA Pathfinder (ASKAP) \citep{Johnston2008} and  the APERture Tile In Focus (Apertif) \citep{Oosterloo2009}. The Wide field ASKAP L-band Legacy All-sky Blind surveY \citep[WALLABY;][]{Koribalski2012}  and the Westerbork Northern Sky \HI\  Survey (WNSHS) will provide an all-sky \HI\ survey at $\sim 15\arcsec-30$\arcsec\ resolution,  a significant improvement upon previous single-dish surveys. WALLABY and WNSHS are expected to resolve thousands of galaxies to an extent  that their kinematical parameters can be retrieved \citep{Duffy2012}. This paper is the first in a series that presents an algorithm for automatically extracting kinematic parameters for these detections.\\
\indent Owing to the fact that in general the orbits of the neutral gas in disc galaxies deviate only slightly from circularity, the common approach to parameterise this geometry and kinematics is to make use of the tilted-ring model \citep{Rogstad1974}. In this framework, gas discs are modelled as a sequence of rings of increasing radius, where each ring has its own geometry and kinematics.  Historically, the fitting was performed on velocity fields constructed from the \HI\ emission line observations.  When analysing velocity fields, the tilted-ring model is typically under-constrained, such that the fitting process needs to be guided interactively (introducing, often subjective, priors to effectively regularise the fitting process). In a manifold of \HI\ studies a process is followed where the velocity field of a galaxy is fitted with the {\sc gipsy} task \rc\ \citep{Begeman1987,vdHulst1992} in several iterations. A typical fit starts by varying several parameters simultaneously to obtain initial guesses, and these parameters are subsequently fitted one at a time until they converge to a solution \citep{Begeman1987,deBlok2008}.\\
\indent This approach has several problems. As the velocity field is used for constructing the model, beam smearing and projection effects can severely affect the final outcome. Already early on it was recognised that the inner rotation curve slope is biased low, due to beam smearing, when the data have less than 7 resolution elements inside the Holmberg radius  \citep{Bosma1978}. At high inclinations the projection effects become so severe that reliable rotation curves can no longer be retrieved from velocity fields \citep{Sancisi1979,Sofue2001}. Even though this effect has been long known, an accurate investigation into the exact inclination at which projection effects become limiting to accurate analysis has not been undertaken.\\
\indent The coupling between the rotational velocities and inclination can also affect tilted ring fits of velocity fields. As we only observe velocities along the line-of-sight, the observed velocities relate to the rotational velocities as v$_{\rm obs}=v_{\rm rot}\times {\rm sin(}i)$. In particular, inclination and rotation velocity become mathematically degenerate in velocity fields of galaxies with solid-body rotation curves, as well as at low inclinations where the line-of-sight component of the rotation is similar in amplitude to the disc velocity dispersion. As a result, galaxies with inclinations
$i<40$\degr are typically not well-suited for tilted ring modelling employing velocity fields (e.g. \citealt{Begeman1987}, but see \citealt{Bershady2010}).\\
\indent  In recent years multiple codes have become available that are optimised for different aspects of  2D kinematic analysis. For example, the tilted-ring model has been adapted for integral field unit (IFU) observations of elliptical galaxies at optical wavelengths \citep[{\sc kinemetry},][]{Krajnovic2006}. This technique deals with the underconstraint  by first applying a grid based approach to find the global minima and subsequently applies a $\chi^2$ minimisation. Another technique is to break the degeneracy outright by fitting flat disc models to the data, an approach adopted by \df\footnote{\url{http://www.physics.rutgers.edu/~spekkens/diskfit/}} \citep{Spekkens2007}.\\
\indent \HI\ data as obtained with modern interferometers are mostly 3D, with two spatial axes and one spectroscopic axis. However, when constructing velocity fields  the observations are compressed into a 2D representation of the velocities which can cause a loss of information. Additionally, the method of compression, can add another bias to the data analysis \citep[e.g.][]{deBlok2008}. By fitting the 3D observations directly several problems, such as beam smearing and projection effects, can be overcome. It is also possible to fit instances where the disc crosses the line of sight multiple times. The most extreme cases of the latter example are edge-on galaxies where the disc crosses the line-of-sight an  infinite number of times. While rotation curves alone may be extracted using other methods \citep{Sofue1996, Kregel2004}, 3D fitting techniques are most suitable for extracting kinematical information from such galaxies.  \\
\indent 3D studies can not only negate projection effects but they can also be used to better understand the structure of galaxies \citep[e.g,][]{Voigtlander2013}. The disadvantages of 3D fitting are that the fitting process is computationally expensive and more sensitive to inhomogeneities in the gas distribution \citep{Jozsa2007}. A model describing the full 3D data requires additional parameters which in turn need to be determined or fixed, therefore the models may still be under-constrained and an interactive fitting procedure has so far been required. \\
\indent The Tilted Ring Fitting Code  \citep[\tiri][]{Jozsa2007}\footnote{\url{http://gigjozsa.github.io/tirific/}} has long been the only published and freely available 3D tilted-ring fitting code. The current version of \tiri\  has an unprecedented degree of flexibility, owing to the fact that past development has been focused on enabling detailed multi-layer modelling. Models employing a large range of parameters are required for the interpretation of ultra-deep \HI\ observations of nearby galaxies, such as those in the Hydrogen Accretion in LOcal GAlaxieS (HALOGAS) Survey \citep{Heald2011, Zschaechner2011, Gentile2013, Kamphuis2013,deBlok2014}. Due to the complexity of these gas distributions and the wide range of \tiri\ model parameters allowed, \tiri\ requires interactive fitting and little emphasis has been put on developing a truly automated fitting method. Here, we present our recent development to overcome this weakness of \tiri\ in the limit where all model parameters are directly determined from the data and no priors are used.\\
\indent In this paper, we present and test an IDL code that utilises the strengths of \tiri\ without the need for interaction during the fitting processes. We  test the code on a set of 52 artificial galaxies and  25 real \HI\ observations selected from LVHIS. In addition we explore the differences between the 3D and 2D approaches and different methods of stabilising the fits by comparing the 3D fits with \rc\ and \df\ models. \FAT\ does not utilise all of \tiri's flexibility but aims to reliably fit an asymmetric warp, surface brightness profile and a symmetric rotation curve to a range of different galaxies. As such its output can be used to retrieve these kinematical parameters for large surveys or it can be used as an objective starting point for more complex models. This code is a prototype for the 3D branch of the WALLABY kinematical pipeline. \\
\indent This paper is structured as follows. In $\S$ \ref{pipeline} we detail the philosophy for this pipeline and specify the initial input and structure of the code presented in this paper.  $\S$ \ref{testmodules} introduces  the test modules we have created for the code  and $\S$ \ref{results} presents the results of these tests. Finally we will discuss these results and suggest possible improvements to the code in $\S$ \ref{discussion} and conclude and summarise in $\S$ \ref{summary}.   
\section{A Kinematic Modelling Pipeline}\label{pipeline}
\begin{figure}  
   \centering
   \includegraphics[width=8.5 cm]{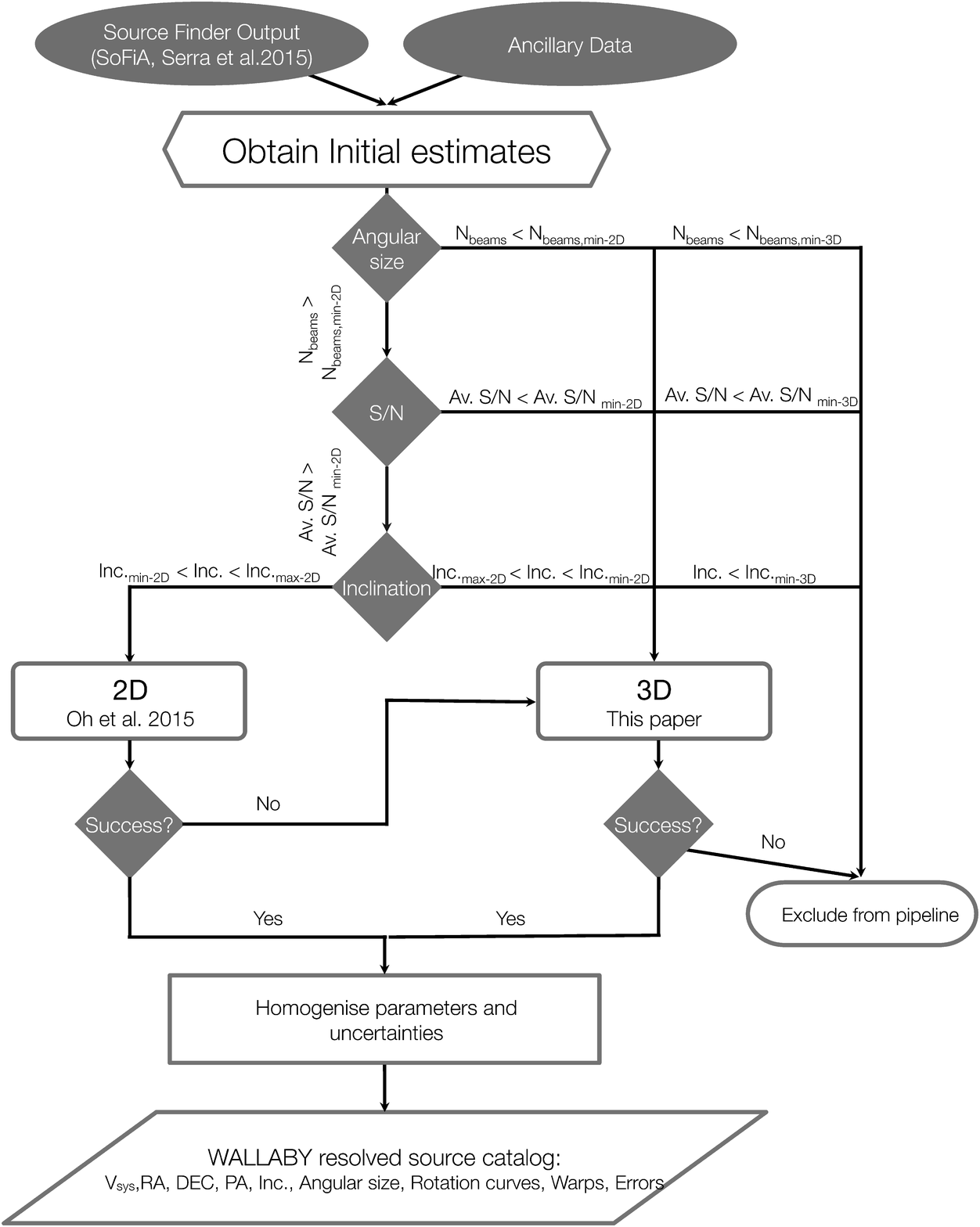} 
   \caption{Conceptual flow diagram of the complete WALLABY kinematic modelling pipeline.}
   \label{fig:flowdiagram}
\end{figure}

\begin{figure}
   \centering
   \includegraphics[width=8 cm]{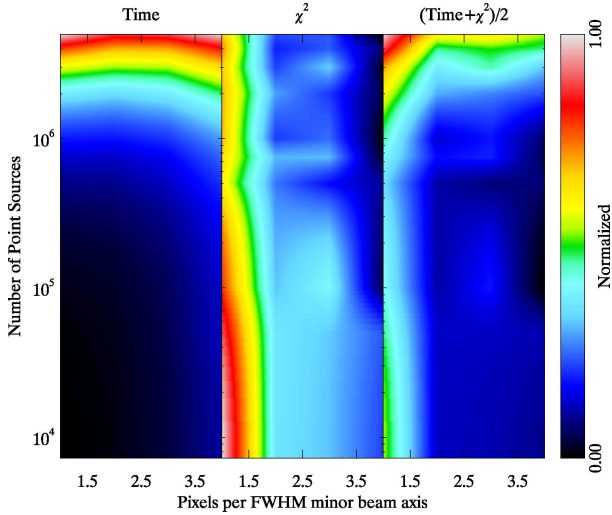} 
   \caption{Comparison of speed vs. accuracy. From left to right panels show duration, $\chi^2$, and their average as determined from a simple fit of five galaxies created with \tiri\ on a grid of pixels per minor axis beam and number of point sources per model. The images are smoothed for display purposes.}
   \label{benchtest}
\end{figure}

Figure \ref{fig:flowdiagram} shows a conceptual flow diagram for the pipeline that is being developed to model the kinematics of resolved detections from  WALLABY and WNSHS. The key principle of this pipeline is that it should automatically extract kinematical parameters in a uniform and reliable manner. This paper will focus on the 3D branch of the pipeline, whereas the 2D branch will be presented in an accompanying paper \citep{Oh2015}.\\
\indent The input for the pipeline can be either a velocity field or a data cube, depending on the initial estimates that are extracted from the data through the source finder. Based on these estimates the pipeline will perform (more efficient) 2D fitting in cases where this approach is thought to be reliable and (more direct) 3D fitting for more complicated cases. If the 2D fit fails the pipeline will try to fit the full data cube in the 3D branch. \\ 
\indent The 3D branch presented in this paper is Fully Automated \tiri\  (\FAT) which is currently an IDL  wrapper around \tiri\ \citep{Jozsa2007}.  
The aim of \FAT\ is to extract from an \HI\ line observation the surface brightness profile (SBR), position angle (PA), inclination (INCL) and rotational velocity (V$_{\rm rot}$) as a function of radius and  the scale height (Z0), intrinsic dispersion (SDIS) and central coordinates (RA, DEC, V$_{\rm sys}$) as single values without any human interaction. In the following we describe the pipeline inputs and the fitting procedure.\\
\indent  Due to the continuously improving computational capabilities \FAT\  is designed with the philosophy that accurate results are far more important than speed. Even so, in this section we first investigate the speed bottlenecks in a systematic way. In \tiri\ there are two major ways in which the fitting process can be significantly slowed. The first is having a large amount of voxels in the data cube. The other bottleneck is related to how \tiri\  creates the models. A model is created by populating the model geometry with \HI\ clouds, i.e. point sources, in a Monte-Carlo fashion and subsequently convolving this point source model with a 3D-Gaussian \citep{Jozsa2007}. The user can control the number of point sources used in the model with a specific input parameter that assigns an approximate flux to each point source. A large number of point sources, i.e a low flux for each point source, will result in a large increase in computing time. However, a low number of point sources will increase the discretisation noise and results in lower accuracy.\\
\indent In order to determine the optimum speed settings without loss of accuracy we have tested a single \tiri\ iteration on five artificial galaxies. In this test \tiri\ performs a $\chi^2$ minimisation through sequential fitting of all parameters using the golden-section method \citep{Press1992}. The artificial galaxies were created with \tiri\ and white noise with  random amplitude was added to mimic observing uncertainties. In these fits we varied the number of pixels across the minor axis FWHM of the beam from 1-4 as well as the number of point sources used to generate the model. \tiri\ fits each parameter sequentially until all parameters are fitted in a so-called loop. It then evaluates whether specific convergence criteria are satisfied. If not, it will start again from the first parameter. We timed the duration of every fit and the average duration of a single \tiri\ loop and averaged these two quantities.  The reason for considering both is that in \tiri\ the number of loops required for convergence can change with changing accuracy. This reasoning was confirmed by small differences in the time grids build up from a single loop and a full iteration. However, as this depends on the complexity of the fit/galaxy we consider both.  The fit and loop times are normalised for each galaxy before averaging.\\
\indent Figure \ref{benchtest} shows the grid for, from left to right, normalised duration, normalised final $\chi^2$ of the model and a speed-accuracy grid in which the former two are averaged. The grid consists of a 4 (pixels per beam) by 10 (number of point sources) roster. Formally the optimal settings for this set of galaxies, i.e. the minimum of the combined duration $\chi^2$ grid, is at 4 pixels per beam and 1$\times10^5$ point sources per model. However, it is clear that accurate results with 3 pixels across the minor axis FWHM  of the beam require at least 3$\times10^5$ point sources per model. Additionally we can see from the left most panel in Figure \ref{benchtest} that for models with more than 2$\times10^6$ point sources, the fitting slows down significantly without a large increase in accuracy. \FAT\ therefore adjusts the point source flux such that the number of point sources per model is always between 5$\times10^5$ and 2.2$\times10^6$. Regridding the input cubes is optional but in this paper we always use data with 4 pixels across the minor axis  FWHM.\\
\subsection{\FAT\ Input}
\FAT\  requires an input catalogue identifying a set of galaxies to be fitted, a configuration file and a data cube following the FITS standard for each galaxy. All fitting is done in observed units, e.g. arcsec not kpc, and thus a distance is not required. However, if a galaxy distance is provided outputs are converted to physical units  such as kpc and \msun. Providing a distance can  also help with the initial estimates, especially the initial scale heights and scale lengths as they are based on physical empirical estimates.\\
\indent  \tiri\ requires initial estimates for the kinematic parameters to be fitted. The initial estimates for the velocity width, the size and galaxy centre are obtained from the source finder \sofia \footnote{\url{https://github.com/SoFiA-Admin/SoFiA/}\\ This paper uses v0.4.0} \citep{Serra2015}. \sofia\  also provides a mask that is used for the creation of an integrated moment map and intensity weighted mean velocity field, i.e. moment 0 and moment 1 maps. The reason for choosing \sofia\ is twofold; it provides uniform initial estimates and masks from which to start the fitting process and it is a product of the WALLABY source finding group, thus ensuring compatibility of the pipeline with the WALLABY source finder. \\
\indent The moment maps created from the \sofia\  mask are used to estimate the PA. This estimate is the average of the PA obtained from a simple ellipse fit to the integrated moment map and the PA determined from the angle between the maximum and minimum velocity in the intensity weighted mean velocity field. Following the estimate of the PA, the inclination is  determined as the average of the axis ratio at three different fractions of the peak intensity (i.e. at $\frac{1}{3}\times \rm{I_{peak}}$,  $\frac{1}{2}\times \rm{I_{peak}}$, $\frac{1}{1.5}\times \rm{I_{peak}}$) and five different PAs  as determined from the error on the PA estimate, i.e PA-error, PA-error/2., PA, PA+error/2 and PA+error. As mentioned above, these preparatory steps are fully integrated into the \FAT\ pipeline.
\subsection{The 3D Fitting Procedure}\label{fittingProcedure}
\begin{figure} 
   \centering
   \includegraphics[width=8.5 cm]{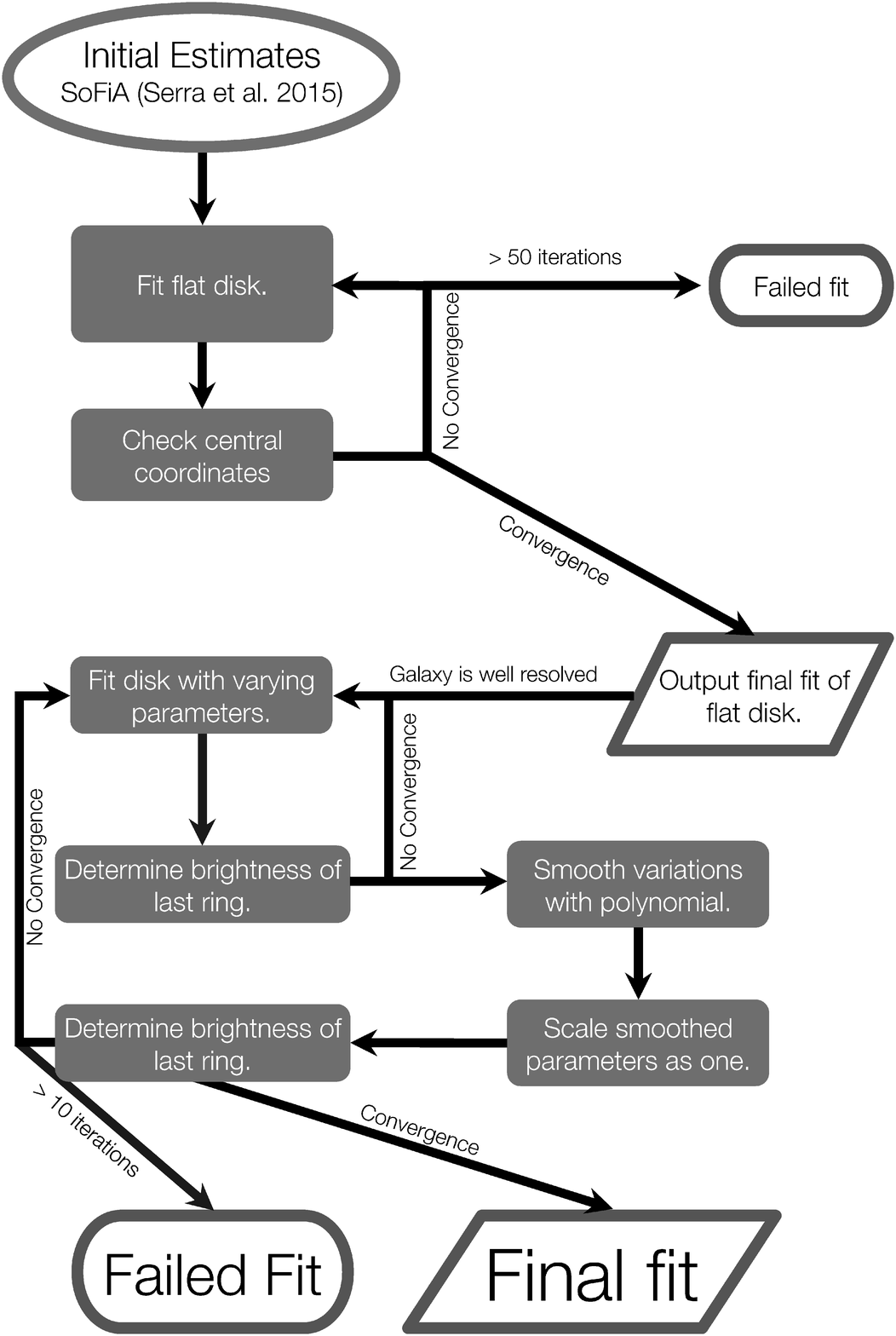} 
   \caption{Flow chart for \FAT. }
   \label{fig:pipelineflow}
\end{figure}

 \FAT\ is a wrapper around \sofia\ and \tiri\ that enables fully automated tilted-ring fits. \tiri\ is run a number of times with increasing model complexity. In addition, \FAT\ adds the following features to the \tiri\ functionality: i) estimating the disk's outer radius, ii) regularising (smoothing) the parametrisation, iii) assessing fit quality/model convergence, iv) controlling fitting speed, v) estimating errors. In the following we describe these steps in detail. While \tiri\ is being constantly developed, most features used in \FAT\ are described in \cite{Jozsa2007}. In addition we utilise \tiri's ability to fit two independent half disks to an observation. The \tiri\ stopping criteria are determined by the SATDELT parameter, which varies throughout the different steps in the code. In the code presented here \tiri\  is always used in golden section mode. \\
\indent The first step in the fitting process (see Figure \ref{fig:pipelineflow}), after initial estimates are obtained, is to fit a flat disc, i.e. a model in which all parameters except the brightness and the rotational velocity  of each ring have the same value at all radii. We adopt the major axis FWHM of the beam as the ring width  in order to produce a stable fit and to ensure that neighbouring rings are independent. The first ring however is merely one fifth of a FWHM in order to keep the central rotational velocity at zero without introducing artefacts. Additionally we halve the ring size for very small galaxies, i.e. less than six beams across the major axis, and double the ring size for large galaxies, i.e. more than 50 beams across the major axis, beyond the inner 20 resolution elements.  \\
\indent In this first step the main aim is to get accurate central coordinates for the model. For the other parameters we aim to fit values that place them within the local minima of their optimal values. We look for convergence on the central position, such that between two subsequent iterations it does not change by more than the size of a voxel. If convergence is not achieved the code will rerun the fitting process with the updated central coordinates. Additionally, \FAT\ checks that the other parameters remain within the boundaries of the initial estimates and updates them when the initial estimates appear incorrect (see Figure \ref{fig:pipelineflow}). If the code fails to converge within 50 iterations, the fitting procedure is halted and flagged as failed.  \\
\indent After a satisfactory fit  for a flat disc is obtained, \FAT\ proceeds to fit radial variations for the inclination, PA, rotational velocities and the surface brightness. In addition to introducing the radial variations we split the model along the minor axis and fit both sides independently except for the rotation curve. For small galaxies ($<$ 6 beams in diameter) we do not vary the inclination and PA radially. \\
\indent In this step the code evaluates several criteria before accepting the fit. Firstly, the internal \tiri\ convergence criteria must be satisfied instead of \tiri\ finishing because the maximum of ten loops through all parameters was reached. If the latter is the case the code will average the radially varying values and try again. Secondly, there must be no more than two consecutive rings close to the boundaries determined in the initial estimates. If there are the boundaries are expanded and \tiri\ is run again. Finally the SBR of the last ring must be in an acceptable range. This is determined by comparing the SBR in the last ring to a cutoff value derived using Equation 4 from \cite{Jozsa2007}:
\begin{equation}
\frac{\sigma_{\tiri}}{{\rm Jy\hspace*{0.1 cm} km\hspace*{0.1 cm} s^{-1}\hspace*{0.1 cm} arcsec^{-2}}}=9.0 \times 10^{-4}  \left(\frac{\Omega_{{\rm ring}}}{\Omega_{{\rm beam}}}\right)^{-0.82}
\end{equation}
 where $\sigma_{\tiri}$ is the cutoff value for each ring, $\Omega_{{\rm beam}}$ is the beam solid angle and $\Omega_{{\rm ring}}$ is the solid angle of the ring\footnote{Note that in \cite{Jozsa2007} the exponent appears to be solely applied to the ring solid angle however this is an error.}. This formula is independent of the noise value and therefore we scale it by the ratio of the noise in the fitted cube to that in the test cubes of 3.6 mJy beam$^{-1}$ \citep{Jozsa2007}. We consider the model to be converged when the surface brightness of the last ring is between 1-5 times this cutoff value. If it is less than the cutoff value the last ring is dropped, whereas if it is more than 5 times the cutoff value the model is extended by one ring. Whenever these iterations lead to a number of rings that has been fitted before, for example by adding a ring and subtracting a ring in the following iteration, or differs by more than 2 rings from the initial estimate we fix the number of rings to remain at this number.\\
\indent Once the previously stated criteria are satisfied we smooth the parameters with a polynomial of order 0, 2, 3, 4 or 5. This is because the simple ring to ring fitting routine of \tiri\ can cause the parameters to vary artificially from ring to ring in a saw tooth pattern and this leaves the possibility for large outliers. The 1st order is excluded because a linear increase from centre to outer radius is not a physical representation of PA and INCL. To determine the order of the polynomial to use for each parameter we fit, through a Monte Carlo approach, all polynomials and adopt the one with the lowest reduced $\chi^2$ value. After the smoothing the parameters are refitted, with the ring to ring variations kept constant, and once more the brightness of the last ring is compared to the cutoff value.\\
\indent As a first empirical attempt at estimating errors on tilted ring models, \FAT\ calculates errors for the fitted parameters on a ring by ring basis. These errors are estimated by the maximum of either the variation in the Monte Carlo fits or the difference between the smoothed and unsmoothed values. We want to stress that this is a first empirical attempt at determining errors and is by no means a formal or statistically correct error. Part of the investigation here is to see whether such errors perform reasonably.\\ 
\section{Test Modules}\label{testmodules}
\subsection{Artificial Galaxies}
\begin{figure*} 
   \centering
   \includegraphics[width=16 cm]{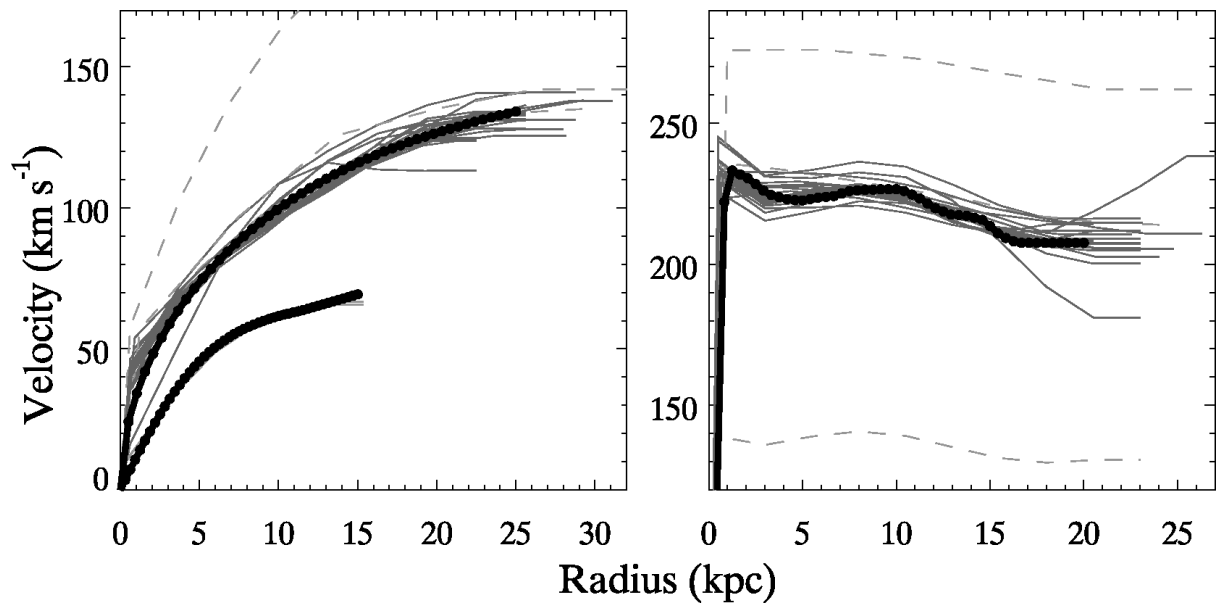} 
   \caption{Rotation curves used and fitted in the artificial galaxy database.  Left panel: dwarf galaxy and intermediate galaxy.  Right panel: massive galaxy. The black circles and lines indicate the input rotation curves, grey solid lines are the reliable output curves, grey dashed lines are from artificial galaxies outside the reliable fitting range (see section \ref{AG}).}  
   \label{fakerotcurs}
\end{figure*}

We have constructed 52 artificial galaxies to test the accuracy of \FAT. We build the models around two different rotation curves (see Figure \ref{fakerotcurs}): one resembling a slowly rising rotation curve typical of intermediate-mass  spiral galaxies and one based on a smoothed version of the rotation curve of NGC 891 \citep{Oosterloo2007,Fraternali2011} in order to represent massive spiral galaxies. For each type of galaxy we adopt an exponential surface brightness profile with a scale length of 10 kpc, which then gets locally perturbed to mimic surface brightness variations. This profile is truncated at a specific radius by decreasing the scale length of the profile by a factor of 20. The values of other parameters of the base galaxies, such as inclination and size, were set to optimal values for tilted ring fitting.  The final model observations are then derived from these base models by varying  one parameter. As we have set the base galaxy values optimally, the effect of the parameter that is varied can be isolated.  The values for the base galaxies are shown in columns two and three of Table \ref{table:Database} and the values to which they were varied are in the fourth column.  We create these artificial galaxies as a distribution of  point sources, add white noise and subsequently smooth them to produce a range in the number of beams across the major axis. \\
\indent In these models the FWHM of the beam is kept constant at 30\arcsec and the physical truncation radius is kept at  25  and 20 kpc  for the  intermediate and massive  galaxy, respectively. In order to vary the number of beams that span the diameter defined by the truncation radius the distance is  adjusted. Afterwards this distance is used to convert the scale heights from fixed physical values to angular scale heights. \\
\indent The noise is calculated as a fraction of the average signal in the noise-free cube; first in each channel the average is calculated of all pixels that exceed 10\% of the maximum signal in the cube. The average of all these averages is subsequently taken as the average signal in the cube.\\
\indent In order to warp some of the test galaxies we vary the angular momentum vector of the initial disc linearly with radius,  in a random direction. Table \ref{table:Database}  lists, in the third column, the variations of $\theta$, the angle between the line-of-sight and the angular momentum vector, and $\phi$, the angle between the North-South axis of the sky and the angular momentum vector. We keep the inner disc flat and start applying this variation to the angular momentum  vector at $\frac{2}{3}$ of the disc extent  such that the maximum variation, as stated in Table \ref{table:Database}, is reached at the cutoff radius. We then project the warped rings  back to an inclination and PA for each ring. This leads to realistic warps with changes in PA and inclination of $\sim$10\degr.\\ 
\indent In total, we produce 25 models for the high mass and intermediate mass galaxy. To study the effect of a reduced velocity resolution (in terms of resolution elements) when fitting dwarf galaxies, we add to the 50 models two further models resembling a dwarf galaxy (see Figure \ref{fakerotcurs}) with the parameters of the base galaxies. The exception to this is  the surface brightness profile, which is produced anew with a scale length of 7.5 kpc and is truncated at a physical radius of 15 kpc. \\
\begin{table*}
   \centering
  
   \begin{tabular}{@{} lll | l  @{}} 
   \hline
    &  \multicolumn{2}{c}{Database Galaxies} & Variations \\
\hline 
      Parameter   &Base 1& Base 2& Sample Values \\
      \hline
         Beams across the major axis  & 16 &16 & 4, 6, 7, 8, 9, 10, 11, 12 \\
         Inclination ($^\circ$)& 60 & 50 &   10, 20, 30, 40, 85, 88, 90 \\
         PA  ($^\circ$)& 45 & 55 & - \\
         Inner-Outer Dispersion (km s$^{-1}$) & 8.0-8.0& 8.0-8.0 & 13.0-7.5\\
         Inner-Outer $\Delta$ $\theta$  ($^\circ$) &0-0& 0-0 & 0-0, 0-9, 0-3\\
         Inner-Outer  $\Delta$ $\phi$  ($^\circ$) &0-0& 0-0 & 0-17, 0-3, 0-11\\
         Inner-Outer scale height (kpc)& 0.3-0.3 &0.2-0.2 & 0.2-0.5, 0.5-0.5\\
         Channel width (km s$^{-1}$)& 4 & 4 & -\\
         Signal-to-Noise& 8 & 8 & 2,4,16\\
         Rotation Curve& intermediate & massive & dwarf \\
        	   \hline
     \end{tabular}
   \caption{Values for the two base galaxies (columns 2 \& 3) and the values by which they  are varied (column 4). We change only one parameter at a time while all other parameters for the base galaxies are unchanged.}
   \label{table:Database}
\end{table*}
\subsection{LVHIS Test Module}
\begin{figure*} 
   \centering
   \includegraphics[width=19cm ,angle=0]{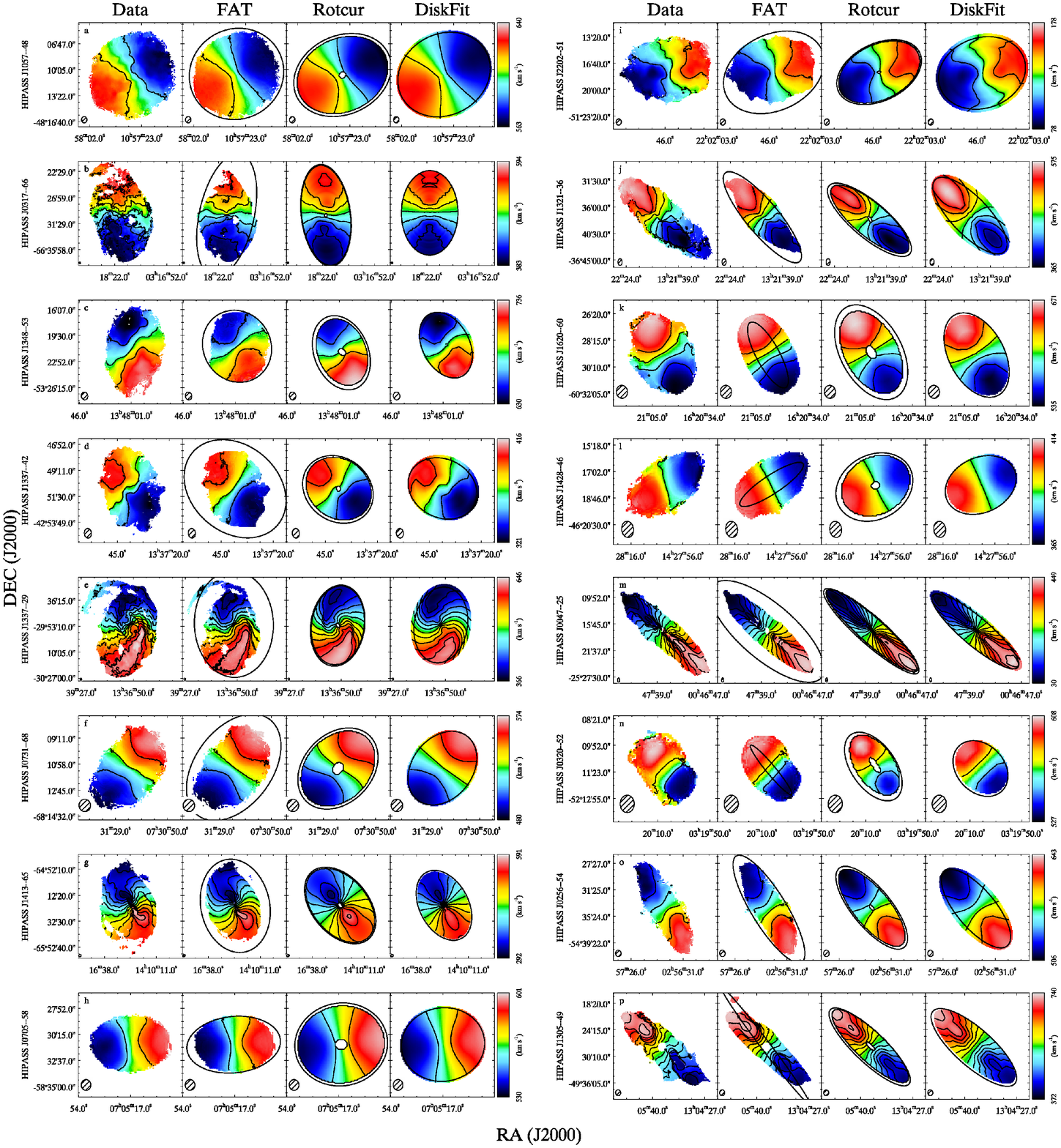} 
   \caption{Velocity fields of the subset of the LVHIS sample that has been fitted with \df, \rc\ and \FAT.  From left to right each row shows the velocity field of the Data as well as the best fitting, \FAT, \rc\ and \df\  models. The first two velocity fields are derived from a Gaussian-Hermite fit.  The \rc\ velocity fields are produced from the fitted parameters with the {\sc gipsy} task {\sc velfi} and the \df\ velocity fields are output from that code. The black contours start at the minimum velocity in the field and have increments of 25 km s$^{-1}$. The black ellipses show the PA and inclination at the furthest extent of the models.}
   \label{mom1}
\end{figure*}
\begin{table*}
   \centering
   \begin{tabular}{@{} lllllccccc @{}} 
   \hline
      \multicolumn{10}{c}{LVHIS Galaxies} \\
\hline 
      HIPASS Name    & Alternative Name &  $\alpha$ & $\delta$ & v$_{\rm sys}$ & W$_{50}$ & PA & Inclination &Distance& \df\ \\
           &  & J2000 & J2000 & km s$^{-1}$ &  km s$^{-1}$ & $^\circ$ & $^\circ$ &Mpc & converged\\
  \hline
HIPASS J0047-25&NGC 253         &00h 47m 31s &-25d 17m 22s & 243$\pm$ 2&407&52 &83$\pm$0&3.1& Y\\ 
HIPASS J0256-54&ESO 154-G023     &02h 56m 55s &-54d 34m 58s & 574$\pm$ 2&122&39 & $\ge 88$&6.8& Y\\  
HIPASS J0317-66&NGC 1313         &03h 17m 57s &-66d 33m 30s & 470$\pm$ 2&168& &36$\pm$7&4.0& Y\\      
HIPASS J0320-52$^*$&NGC 1311         &03h 20m 05s &-52d 11m 34s & 568$\pm$ 5& 80&40 &79$\pm$1&5.3& Y\\ 
HIPASS J0333-50$^*$&IC 1959          &03h 33m 15s &-50d 25m 17s & 640$\pm$ 4&128&147 &88$\pm$4&8.2& N\\ 
HIPASS J0705-58&AM 0704-582      &07h 05m 18s &-58d 31m 19s & 564$\pm$ 2& 68& &65$^{1}$&4.9& Y\\   
HIPASS J0731-68&ESO  59-G001     &07h 31m 20s &-68d 11m 19s & 530$\pm$ 3& 82& &20$\pm$18&4.5& Y\\      
HIPASS J1047-38$^*$&ESO 318-G013     &10h 47m 39s &-38d 51m 45s & 711$\pm$ 7& 42&75 & $\ge 88$& & N\\	       
HIPASS J1057-48&ESO 215-G?009    &10h 57m 32s &-48d 11m 02s & 598$\pm$ 2& 67&72 &64$\pm$27&5.3& Y\\
HIPASS J1219-79$^*$&IC 3104          &12h 19m 04s &-79d 42m 55s & 429$\pm$ 4& 40&45 &59$\pm$6&2.6& N\\ 
HIPASS J1305-40$^*$&CEN 06           &13h 05m 02s &-40d 06m 30s & 617$\pm$ 4& 33& &60$^{1}$&6.1& N\\   
HIPASS J1305-49&NGC 4945         &13h 05m 24s &-49d 29m 35s & 563$\pm$ 3&361&43 &85$\pm$4&4.1& Y\\ 
HIPASS J1321-36&NGC 5102         &13h 21m 55s &-36d 38m 03s & 468$\pm$ 2&200&48 &70$\pm$6&18.8& Y\\
HIPASS J1337-28$^{**}$&ESO 444-G084     &13h 37m 18s &-28d 02m 17s & 587$\pm$ 3& 56& &39$\pm$4&5.1& N\\       
HIPASS J1337-29&NGC 5236         &13h 37m 02s &-29d 55m 03s & 513$\pm$ 2&259& &2$\pm$27&6.8& Y\\       
HIPASS J1337-39$^*$&           &13h 37m 30s &-39d 52m 56s & 492$\pm$ 4& 37& &36$^{1}$&4.8& N\\   
HIPASS J1337-42&NGC 5237         &13h 37m 47s &-42d 50m 51s & 361$\pm$ 4& 77&128 &35$\pm$0&3.7& Y\\
HIPASS J1348-53&ESO 174-G?001    &13h 48m 01s &-53d 21m 31s & 688$\pm$ 3& 71&170 &76$\pm$11& & Y\\ 
HIPASS J1403-41$^{**}$&NGC 5408         &14h 03m 21s &-41d 22m 26s & 506$\pm$ 3& 62&62 &55$\pm$8&4.9& N\\ 
HIPASS J1428-46$^*$&UKS 1424-460     &14h 28m 06s &-46d 18m 32s & 390$\pm$ 2& 48& &73$^{1}$&3.4& Y\\   
HIPASS J1413-65&Circinus        &14h 13m 27s &-65d 18m 46s & 434$\pm$ 3&242&40 &64$\pm$4&4.2& Y\\ 
HIPASS J1441-62$^*$&           &14h 41m 37s &-62d 44m 38s & 672$\pm$ 8& 52& & & & N\\	       
HIPASS J1501-48&ESO 223-G009     &15h 01m 08s &-48d 17m 04s & 588$\pm$ 2& 61&135 &44$\pm$19&6.0& N\\
HIPASS J1620-60$^*$&ESO 137-G018     &16h 20m 56s &-60d 29m 18s & 605$\pm$ 3&139&30 &73$\pm$2&5.9& Y\\ 
HIPASS J2202-51&IC 5152          &22h 02m 41s &-51d 17m 37s & 122$\pm$ 2& 84&100 &51$\pm$4&1.8& Y\\ 
          \hline
     \end{tabular}
   \caption{Overview of the 25 LVHIS galaxies selected for testing the code. All positional data is taken from Table 2 in \citet{Koribalski2004}. The distance is the currently stated average distance in the NASA Extra-galactic Database (NED). The PA and  axis ratios are taken from \citet{Lauberts1982}. The inclination is  calculated from the stated axis ratios \citep[see ][]{Aaronson1980} and the error is the difference between the value calculated from the axis ratios in \citet{Lauberts1982} and those provided in NED. If no error is stated this means that the values were the same or solely from NED (marked  $^1$). $^*$Galaxies with less than eight beams across the major axis in \FAT\ fits. $^{**}$The \FAT\ failed for these galaxies. }
   \label{LVHISTable}
\end{table*}
\nocite{Koribalski2004}
\nocite{Lauberts1982}
Real galaxies can contain many deviations from highly symmetrized artificial galaxies.  We therefore run \FAT\ on real observational data, comparing it to the 2D fitting methods \rc\ and \df. Since we do not know the correct solution a priori, the comparison mainly aims at finding systematic differences between 2D and 3D methods. \rc\ and \tiri\ fit the same model in different dimensions and both codes need  to be guided by hand to an acceptable result or, as is attempted here, embedded in a wrapper that makes assumptions about the properties of real galaxies and applies them automatically. \\
\indent For \df\ the problem is not under-constrained as it assumes the disc to be flat; this allows the code to fit global minima and to calculate realistic error bars. \df\ can include streaming motions into the final model to simulate bars or radial motion and also includes a model for a weak, quadratic, symmetric warp. Therefore, the comparison between \rc\ and \df\ can be used to investigate systematic errors introduced by complex warps in the outer parts and bars in the inner parts of these galaxies. \\
\indent In order to test the pipeline's 2D and 3D components we selected 25 galaxies from the southern sample of the LVHIS catalogue \citep{Koribalski2010}. LVHIS  is a complete sample of galaxies selected from HIPASS \citep{Koribalski2004,Meyer2004} with a systemic velocity $\le 550$ km s$^{-1}$ ($\sim 10$ Mpc). The southern sample was further restricted to have  $\delta<-30$\degr. This sample was observed with the Australia Telescope Compact Array (ATCA) with similar angular and velocity resolution to the proposed values for WALLABY (FWHM $\sim$30\arcsec, channel width $\sim$4 km s$^{-1}$). \\
\indent The 25 galaxies were selected by taking the LVHIS \HI\ mass function and selecting the first two most regularly rotating galaxies, as determined from a visual inspection of the velocity fields, from 0.1 dex-wide mass bins. This  sample was interactively fitted with \rc\  \citep[for a standard description of the approach see][]{Oh2008}. In particular, it was decided by inspection whether or not the galaxy kinematics were sufficiently well represented by a rotating disc or whether radial variations of PA and INCL were required. All the \rc\ models are circularly symmetric.\\
\indent  Subsequently the fits  were used as an input  for \df\ \citep{Spekkens2007} in a semi-automated routine, in order to avoid a bias  the \rc\ fit outputs were varied randomly by up to 5$^\circ$ in PA and inclination and 10 km s$^{-1}$ for the systemic velocity to obtain initial guesses for \df. Each of the galaxies was then fitted with a rotation-only model using \df. When the residual velocity fields for these fits showed clear indications of noncircular motions, we added bi-symmetric flows or a quadratic warp to the models.  For \df\ we excluded galaxies that are too small or that have too low rotation velocities. The models used in this discussion all converged. \\
\indent From the initial 25 galaxies we fitted 16 reliably with \df\  and 25 with \rc. Table \ref{LVHISTable} shows the basic parameters of these 25 galaxies as found in the literature. The last column indicates wether the galaxy was successfully fitted with \df. When fitting with \df\ and \rc\  we used velocity fields derived from the data cubes by fitting a Gauss-Hermite polynomial to each line profile (see Figure \ref{mom1}, left panels; only galaxies that are fitted with all three codes are displayed).
\section{Results}\label{results}
\subsection{Artificial Galaxies}\label{AG}
\begin{figure*} 
   \centering
   \includegraphics[width=16 cm]{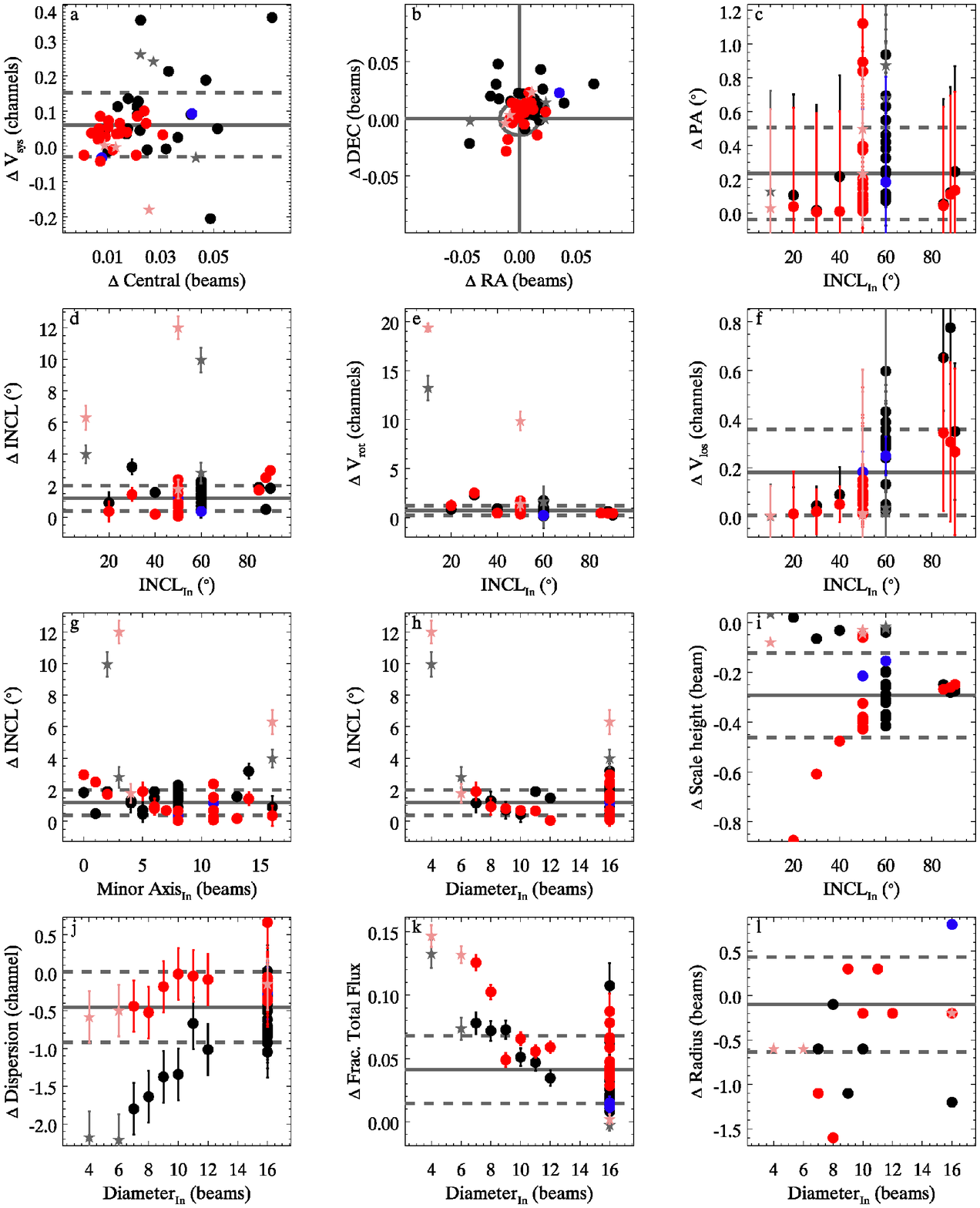} 
   \caption{Difference between the artificial galaxy inputs and  \FAT\ output. On the y-axis the difference between the model input and  best fitting \FAT\ values is shown for various parameters. The solid line shows the mean deviation over the whole sample and the dashed lines show the 1$\sigma$ offsets from this average. The errors on the points are calculated as the average of the \FAT\ error for all rings  divided by $\sqrt{{\rm N}}$, where N is the number of rings in the model. The stars show galaxies outside the reliable range for \FAT\ whereas circles are inside this range. The colour coding indicates: blue - dwarf galaxy, black - intermediate galaxy, red - massive galaxy.The panels show a)  central coordinates difference vs. systemic velocity difference, b) declination difference vs. right ascension difference, c) PA difference vs. inclination, d) inclination difference vs. inclination, e) rotational velocity difference vs. inclination, f) velocity along the line of sight vs. Inclination, g) inclination difference vs. the number of beams across the minor axis, h) inclination difference vs. maximum diameter, i) scale height difference vs. inclination, j) dispersion difference vs. maximum diameter, k) surface brightness difference vs. noise, and  l) model extent difference vs. model extent. Panels c), d), e), f), g) and h) show absolute differences (see text). }
   \label{ArtifOverview}
\end{figure*}
The results of the test on artificial galaxies are in Figure \ref{ArtifOverview} which shows the difference between input and the best-fitting values returned by \FAT. This difference is calculated either as the simple difference for  parameters that do not change with radius, e.g. the dispersion or systemic velocity, or as the error-weighted absolute mean deviation for parameters that vary with radius, i.e. PA, INCL, SBR and VROT. While the former is expected to oscillate about a value of zero across the sample as a whole, the latter is intrinsically a positive quantity. The star symbols show artificial galaxies that lie outside the range in which \FAT\ is deemed reliable. This range is determined in an iterative process by identifying galaxies for which a single fit parameter lies more than 3$\sigma$ from the mean absolute deviation over the whole sample. Subsequently these galaxies are excluded when calculating the averages and rms. The following section explains where \FAT\ fails in more detail for each parameter and what the final limits are.\\
\indent We express the results in Figure \ref{ArtifOverview} in units of beam (30\arcsec), channel (4 km s$^{-1}$), noise and beams across the major axis where possible so that they may be scaled to other emission line observations. 
\subsubsection{Central Coordinates \& Position Angle}
We see that for the test galaxies the pipeline recovers the input of most parameters to within the estimated errors.  The central coordinates (Figure \ref{ArtifOverview}, panel a and b) are accurate to a fifth of a channel in systemic velocity and the discrepancy in RA and DEC is in general lower than a tenth of a beam.\\
\indent The PA (Figure \ref{ArtifOverview}, panel c) has a mean absolute deviation from the input values of 0.23$^{\circ}$.  It is clear from this panel that this deviation is  in general much smaller than the estimated errors and hence we deem the PA to be fitted successfully for all artificial galaxies. \\
\subsubsection{Rotation and Inclination}
The second row from the top of Figure \ref{ArtifOverview} shows the results for the inclination (panel d), rotation (panel e) and the line of sight velocity (v$_{\rm rot}\times {\rm sin(}i)$) (panel f). It is immediately clear from the last panel  (panel f) that the line-of-sight velocities are recovered to within a channel. The trend with inclination is caused by the weighted average of the deviation: at lower inclinations, the range in line-of-sight velocities becomes smaller and smaller and hence the deviations become smaller.\\
\indent The fact that these line-of-sight velocities are well fitted means that the large deviations in rotational velocities (panel e) at low inclination ($< 20\degr$) are caused by the errors in inclination. This is understandable as an error of 1$^{\circ}$ in inclination immediately leads to an offset of $\sim$ 4 km s$^{-1}$ ($\sim$1 channel in this test sample) in rotational velocity for a rotation velocity $\sim$ 200 km s$^{-1}$. The code however amplifies these  errors,  and it is not clear why.\\
\indent There are four galaxies with rotation curves that deviate from the actual values by more than 3$\sigma$ when considering the rotation curve. Clearly the fit is no longer reliable at 10\degr\ inclination. The outlier at 50\degr\ is an artificial galaxy with only 4 beams across the major axis. This is also visible in Figure \ref{fakerotcurs} where the two dashed rotation curves for the massive galaxy represent these galaxies. The fourth outlying galaxy is  the massive galaxy at 30\degr\ inclination; it has a deviation that is a tenth of a channel more than 3$\sigma$ from the mean deviation ($\sigma=0.5$ channel and the mean deviation=0.7 channel) when the error (0.2 channel) is included. It seems that this galaxy falls victim to the success in the other galaxies and is only an outlier due to the very small scatter in the other fits. Since the galaxies at $i=$20\degr\ and the intermediate galaxy at $i=$30\degr\ are fitted well we do not exclude this outlier, hence it appears as a red dot rather than a star in Figure \ref{ArtifOverview}. From this test we conclude that \FAT\ can reliably fit rotation curves in a range from 20\degr-90\degr\ inclination.\\
\indent This lower limit in inclination implies \FAT\ can model galaxies that are more face on than 2D methods can  \citep[40\degr; ][]{Begeman1987}. It is possible that \FAT\ can fit galaxies at even lower inclinations reliably; however, our simulations do not probe much below this threshold and therefore we adopt 20\degr\ as a conservative lower limit.\\
\indent We also highlight that, while the errors at low inclinations are high compared to what is reachable for higher inclinations, they might still be statistically unbiased, as panel e) (line-of-sight velocity) suggests. Another test series with a significantly larger sample size is required to determine if this is indeed the case.\\
\indent In inclination  (Figure \ref{ArtifOverview}, panel d) the typical error is $\sim$ 1\degr\ and the scatter in accepted fits is 0.8\degr. The two outliers at inclinations of 50\degr\ and 60\degr are galaxies with 4 beams across the major axis. This is also visible in panel h of Figure \ref{ArtifOverview} which shows the difference in inclination plotted against the number of beams across the major axis. From this plot it is immediately clear that, even in 3D, the inclination estimate is hampered  by the number of beams across the major axis. There is a clear trend of increasing absolute deviation and error increase with a decreasing number of beams across the major axis. Even though the deviations are both positive and negative in the sample as a whole, for these smaller galaxies the deviations are all positive. \\
\indent This behaviour also affects the rotation curve but not as strongly as might be expected. Only in the case of the massive galaxy with four beams across the major axis does the deviation from the input stand out, i.e. the outlier in panel e at 50\degr\ inclination. However, an inspection of the rotation curves in these galaxies shows that in general the shape of the inclination and rotation curve are retrieved correctly, albeit with an offset. Only at 4 beams across the major axis does the fit fail completely, returning an inclination that is far offset from the input value. The outliers at low inclination were already deemed unreliable from their deviations in rotational velocity. \\
\indent  Figure \ref{fakerotcurs} shows that the best fitting rotation curves of two massive galaxies have significant deviations at large radii. The curve with an outer decrease corresponds to a warped artificial galaxy. Here the variation in inclination cuts through zero and subsequently increases again beyond the cutoff radius. This causes the inclinations in the outer radii to be overestimated thus lowering the rotation curve amplitude. The curve showing the large outer increase corresponds to a massive artificial galaxy with 8 beams across the major axis. This behaviour is caused by the fit being extended beyond the truncation radius, with the consequence of an arbitrary fitted value for the rotation velocity. It is not clear why this happens in this galaxy but not others, but clearly panel l in Figure \ref{ArtifOverview} shows that this is the most extreme case.\\
\indent  An individual inspection of the fits showed that at high inclination the fits for the massive galaxy are not as good as their overall deviations indicate. The galaxies at 88\degr\ and 90\degr\ have inner inclinations which are underestimated by $\sim$3\degr-5\degr. This does not happen for the intermediate galaxies; therefore it is likely related to the shape of the rotation curve. This clearly points to difficulties with fitting of inner parts of edge-on galaxies which is not unexpected. This behaviour should be reflected by an increase in the error towards the inner parts of these highly inclined galaxies. However, the current version of \FAT\ fails to pick this up. As this does not affect any other aspects of the fit and the deviations are still acceptable we do not exclude this range. Even so, we stress that care should be taken when \FAT\ is applied to highly inclined (INCL $\ge 88\degr$) galaxies.\\
\indent Panel g in Figure \ref{ArtifOverview} shows the difference in inclination plotted against  the number of beams across the minor axis. This panel makes it clear that in these 3D fits the errors are truly driven by inclination or  the number of beams across the major axis and not by the number of beams across the minor axis.
\subsubsection{Scale Height}
Panel i of Figure \ref{ArtifOverview} shows that on average the scale height is over estimated by roughly a third of the beam. The typical distance for this test set of galaxies is $\sim$ 19 Mpc. This means that the scale heights of the models are about a 10th of the beam. The fact that the problem is worse for the massive galaxy, which has a smaller intrinsic scale height, points also to the scale height being unresolved. However, the trend of increasing deviations  towards lower inclination points to the scale height being less constrained at lower inclinations. This is expected, since the most important effect of an increased scale height, a thickening of the projected disc along the minor axis, scales with the sine of the inclination.\\
\indent In order to determine the exact cause of this error a larger database more specific to this problem is required. A comparison with the un-regularised fits hints, at least at low inclination, to the various asymmetries and noise peaks being compensated in this parameter as the discrepancies between fit and model are much lower in the un-regularised cases. Like the inclination, we notice the tendency to stretch the galaxy model along the minor axis. We hence conclude that scale heights less than 1/10th of a beam are not fitted accurately.\\
\subsubsection{Dispersion}
The fitted dispersion is in general overestimated by half a channel (panel j in Figure  \ref{ArtifOverview}). Even though half a channel is reasonable as the error one can expect, the systematic nature of the offset is remarkable. The trend of increasing deviation with decreasing number of beams across the major axis points towards a beam smearing effect. This idea is strengthened by the fact that the intermediate mass galaxy, which has a rotation curve with strong curvature, is more affected than the massive galaxy with a constant or approximately linear rotation curve. Since \tiri\ models a rotation curve by linearly interpolating the rotation velocity between ring radii, the average deviation of the true rotation curve and the modelled one is much higher for the intermediate mass galaxy than for the high mass galaxy. We suspect that \tiri\ compensates for this by increasing the dispersion more for the intermediate mass galaxy and less for the high mass galaxy.\\
\indent \FAT\ does not provide an error on the dispersion but we estimated an error on each fit of 0.34 channel from the scatter in the fits with 16 beams across the major axis. This means that  the intermediate galaxy with six  beams in diameter has a  deviation in dispersion more than 3$\sigma$ larger than the mean absolute deviation in the sample. As this is clearly a continuous trend we conclude that \FAT\ becomes unreliable for galaxies with less than seven beams across the major axis. This is a conservative limit as clearly this depends on the shape of the rotation curve. \\
\indent At this stage it is unclear whether the continuous overestimation of the dispersion is an artefact of the code, or is related to how the artificial galaxies are constructed, or whether the small deviations in other parameters are compensated by offsets in dispersion. This latter option does not seem to be the case though as the dispersions are  similar for the regularised and unregularised fits.\\
\indent Of additional interest here are the two galaxies where the input dispersion is varied from 13.0 km s$^{-1}$ in the innermost radius  to 7.5 km s$^{-1}$ at the outer most radius. In case of the intermediate mass galaxy the code fits a total dispersion of 12.9 km s$^{-1}$  indicating that the fit is driven by the central regions of the galaxy. However, for the massive galaxy the code fits 10.3 km s$^{-1}$  which is intermediate to the two extremes. This galaxy is also the only galaxy with a positive offset because for single values, such as the dispersion, only the value in the central ring is considered. This result means that from these galaxies it is unclear how the code deals with varying dispersions and a larger test set is required.
\subsubsection{The Surface Brightness \& Signal-to-Noise}
The surface brightness is captured well in general as shown in panel k of Figure \ref{ArtifOverview}. The average deviation is $\sim$ 4\%. This panel shows the difference between the flux in the input model and \FAT\ output, as a fraction of the total flux in the input model, against the number of beams across the major axis. The total flux was measured in the original model cubes without added noise, but with the correct resolution,  in voxels where the flux exceeded the subsequently added noise. The error is the formal error expected from the noise and number of added resolution elements.\\
\indent It is clear from Figure \ref{ArtifOverview} panel k that the number of beams across the major axis severely affects the fitted flux. In this case galaxies that lie more than 3$\sigma$ from the mean are those with four beams across and the massive galaxy with six beams across. These deviations appear to be connected to those in dispersion. However, a larger deviation in dispersion seems to lead to a lower deviation in total flux, or vice versa. At this point it is unclear how these two parameters correlate with each other although the clear increase with number of beams across the major axis points towards a resolution effect.\\
\indent \FAT\ underestimates the total flux in all models. This is partially due to the models being constructed at lower resolution than the artificial galaxies. This means that the \FAT\  models are unable to capture the high flux peaks in the artificial galaxies. From this dataset it is unclear whether this fully explains the offset. However, there is also a small trend with inclination (not shown) where the fits at lower inclinations capture the total flux much better than those at high inclination. This confirms the idea that the offset is mostly caused by the peak flux being underestimated as this ring of high flux would cover a larger sky area in lower inclination galaxies and would thus be easier to fit.\\
\indent Besides the trends with the number of beams across the major axis and inclination the total flux also shows a weak trend with signal-to-noise. The two points with the largest deviation are the galaxies with a signal-to-noise of two. This is expected as the lower signal-to-noise would make it harder to determine the correct flux. The lower signal-to-noise  hides more of the flux in the noise and hence a model with lower total flux is fitted to the input. However, the deviation of these galaxies is still acceptable.\\
\subsubsection{Extent of the models}
The last panel in Figure \ref{ArtifOverview} (panel l) shows the difference in adopted outer radius vs. input outer radius. \FAT\  estimates the size of the galaxies in this test sample quite accurately. Since the edges in the  artificial galaxies are created by decreasing the scale length by a factor of 20 at the cutoff radius, \FAT\ can be used to detect these edges in real galaxies if they exist. However, as at the given edge the emission does not immediately disappear but continues into the noise, \FAT\ sometimes overestimates the outer radius by 1 beam.  This is not unexpected.  There appears to be a trend in which the overestimation increases for galaxies with fewer beams across the major axis.\\

\indent From these tests we conclude that \FAT\ fits galaxies well in the inclination range 20$^{\circ}$-90$^{\circ}$ as long as the errors produced by the pipeline are taken into account. At an inclination of 10\degr\ the errors in rotation velocity become so large that they render the derived rotation curve meaningless. However, considering the derived errors, the true rotation curve is still reproduced accurately in a purely statistical sense. The same is true for the inclination. The minimum size that can be reliably fitted without any priors is a final model with 7 beams across the major axis.  In galaxies that are less resolved the dispersion or surface brightness is seriously  misfitted, most likely to counterbalance the failure of the current \tiri\ implementation to implement curvature, between rings, in a small scale model. However,  because \FAT\ overestimates the size of the galaxy to a larger extent when it is smaller, the models fitted to artificial galaxies with six beams across can have 8 beams across the major axis in the output. As we want to determine reliability from the fitted models we conclude that without additional input only \FAT\ models with eight beams or more across the major axis and between 20$^{\circ}$-90$^{\circ}$ inclination are reliable. 
This is a very conservative limit because  the tests here show that  fits to smaller galaxies (up to six or seven beams across the major axis) can be accurate. \\
\indent It is clear from Figure \ref{ArtifOverview} that \FAT\ is slightly better at fitting the massive galaxy (red) than the intermediate mass galaxy (black). This affects the fitted dispersion the most but is also visible in other parameters. Further tests are required  to see whether this is due to the shape of the rotation curve or because of the higher rotational velocities. The two dwarf galaxies simulated here unfortunately do not shed any further light on this matter.
\subsection{LVHIS Sample}\label{RG}
\begin{figure*} 
   \centering
   \includegraphics[width=15.5 cm]{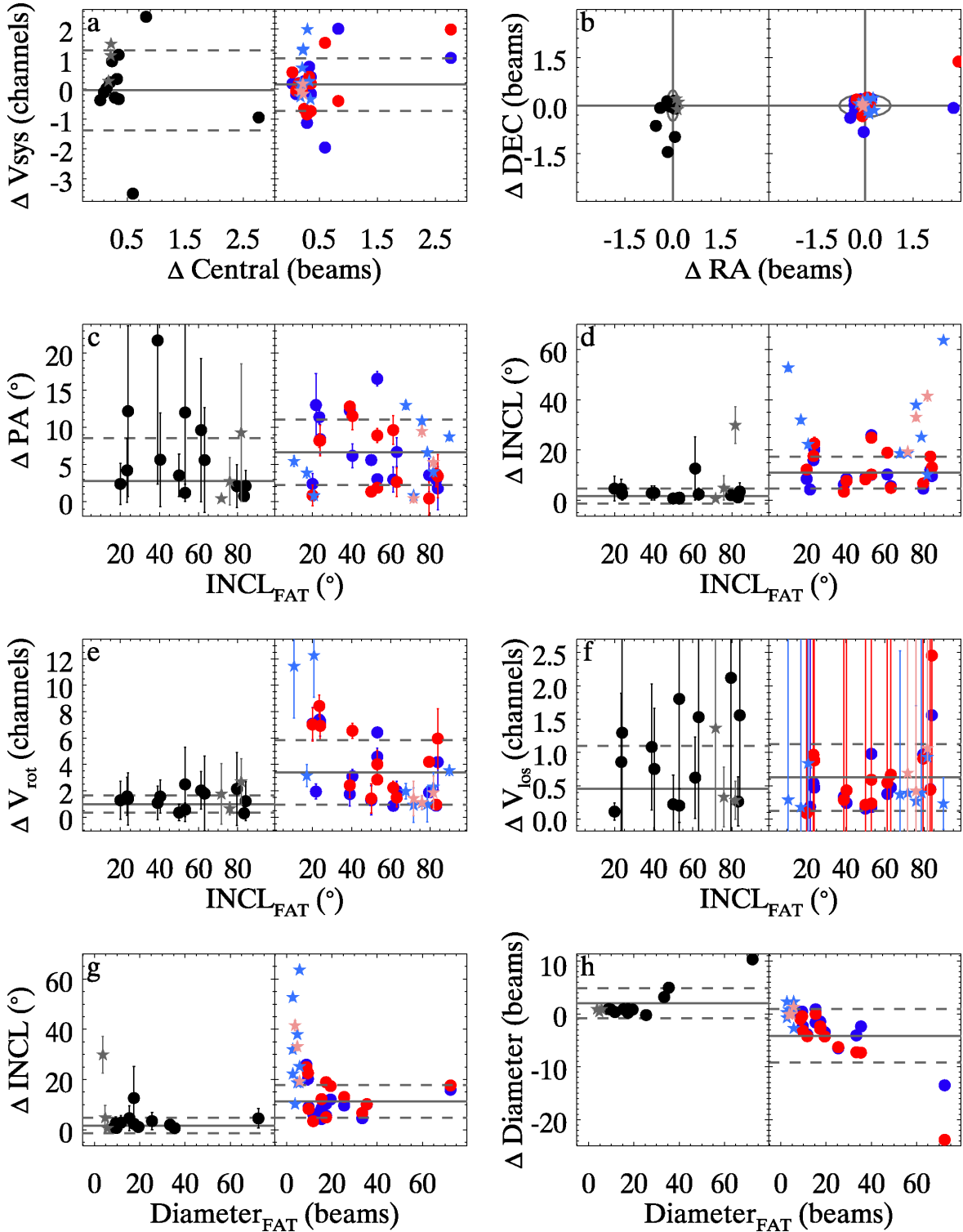} 
   \caption{Differences between the final fits with \FAT, \rc\ and \df. Left hand side of each panel: the difference between \rc\ and \df, i.e. \rc\ - \df, Right hand side of each panel: blue: the difference between \rc\ and \FAT, red: the difference between \df\ and \FAT.  The solid grey lines show the weighted average and  the dashed lines show the  $1\sigma$ deviation from this average in each plot. Stars correspond to final fits outside the previously determined reliable range (see Section \ref{AG}) and circles to ones inside these limits. The errors on the points are calculated as the average of the error for all rings  divided by $\sqrt{{\rm N}}$ with N the number of rings in the model. The panels show a)  Central coordinate difference vs. Systemic velocity difference, b)  Right Ascension difference vs. Declination difference, c) PA difference vs. Inclination, d) Inclination difference vs. Inclination, e) Rotational velocity difference vs. Inclination, f) Velocity along the line of sight vs. Inclination, g) Rotational velocity difference vs. Model extent, h) Model extent difference vs. Model extent. Panels c), d), e), f) and g) show absolute differences (see text).}
   \label{LVHISOverview2}
\end{figure*}
\begin{table*}
   \centering
   \begin{tabular}{@{} lrrrrrr @{}} 
	\hline
Parameter & Mean$_{{\rm RC - DF}}$& $\sigma_{{\rm RC - DF}}$ & Mean$_{{\rm RC - \FAT}}$ & $\sigma_{{\rm RC - \FAT}}$  & Mean$_{{\rm DF - \FAT}}$ & $\sigma_{{\rm DF - \FAT}}$  \\
\hline
RA (beam)& -0.24&   0.19&  0.09&   0.79&  0.22&   0.82\\
DEC (arcsec)& -0.24&   0.48& -0.09&   0.27&  0.13&   0.40\\
V$_{\rm sys}$ (km s$^{-1}$) & -0.02&   1.34&  0.13&   0.92&  0.17&   0.83\\
PA ($^{\circ}$)&  2.75&   5.79&  7.41&   4.53&  5.81&   4.18\\
Inclination ($^{\circ}$)&  1.72&   3.00& 10.47&   5.91& 11.71&   6.59\\
V$_{\rm rot}$ (km s$^{-1}$) &  0.99&   0.65&  3.17&   2.41&  3.66&   2.43\\
V$_{\rm rot}\times$sin($i$) (km s$^{-1}$)&  0.46&   0.64&  0.56&   0.39&  0.70&   0.58\\
Max. Diameter (beam)&  0.96&   2.85& -3.11&   3.56& -5.23&   6.06\\
   \hline
      \end{tabular}
   \caption{The mean absolute deviation and rms scatter of the difference between the classical \rc\ and \df\ fits, the \rc\ and  \FAT\  fits and the  \df\ and \FAT\ fits for galaxies in the reliable range. For the first and last 2 columns this is for 11 galaxies and the middle 2 columns is a sample of 22 galaxies.}
      \label{tab:ParametersLVHIS}
\end{table*}
\tiri\ and  \FAT\ flag  25 of the 25 LVHIS galaxies  as successfully fitted. A visual comparison between the  \FAT\ models and the data cubes confirms that the intensity distributions of  all models, barring two (HIPASS J1337-28 and HIPASS J1403-41), correspond to the data.  For HIPASS J1403-41, the kinematics are complex and not well reproduced by a tilted-ring model \citep[see][]{vEymeren2010}, while for HIPASS J1337-28 imaging artefacts misguide the fitting procedure. These two galaxies are marked with $^{**}$  in Table \ref{LVHISTable} and excluded from further analysis in this section. Optimally, \FAT\ should be able to provide the metrics to exclude such cases. This task will be the subject of future research. \\
\indent In the following we compare the parameters obtained from \FAT\ to those obtained from \rc\ and \df\ fits as well as to the literature in order to get an idea of the errors and differences between the codes.\\
\indent An overview of the results of the fits by \FAT\ for the sample selected from LVHIS is given in Figure \ref{LVHISOverview2} and a subsample of velocity fields for the data  and the three codes is shown in Figure \ref{mom1}. The panels in Figure \ref{LVHISOverview2} show the difference between the \rc\ and \df\ fits (black symbols, left side of a panel), the difference between the \rc\ and \FAT\ fits (blue symbols, right side of a panel), and the difference between the \df\ and \FAT\ fits (red symbols, right side of a panel) on the y-axis plotted against various parameters. As for the artificial galaxies, in cases where the values may vary with radius we calculate an absolute weighted mean deviation of the ring differences and simply take the difference between values that do not vary with radius. As the fits were sometimes performed using different ring sizes, we performed a linear interpolation of the \rc\ or \df\ fit to the \FAT\  radii when needed.\\
\indent  The errors on the points were calculated in the following manner. For parameters that have errors the errors for for both codes were added in quadrature and subsequently averaged and divided by the square root of the number of rings. For parameters that were provided without error no error is shown in Figure \ref{LVHISOverview2}. Table \ref{tab:ParametersLVHIS} shows the average deviation of the total sample and the rms  of the sample that falls within the previously determined reliable \FAT\ fitting range. These values are also indicated as the solid grey lines (average) and the dashed lines (1$\sigma$ deviation from the average) in Figure   \ref{LVHISOverview2}.  Galaxies that fall outside the range determined in section \ref{AG} are marked by star symbols.\\
\indent It is important to note that the \FAT, \rc\ and \df\ models do not always have the same number of degrees of freedom. For example, sometimes the researcher chooses not to include a warp in the \rc\ model where \FAT\ decided to fit a warp; also, it can happen that in the \df\ models streaming motions are included. We compare these final models because we want to investigate how these different choices and methods affect the fits. Additionally, we want to compare these methods at their best and minimise the error on the fits. In the following sections we will discuss the fitted parameters in more detail. \\
\indent Applying the conservative limits derived in section \ref{AG}, thirteen galaxies fitted with \df\ and fourteen galaxies fitted with \rc\ can be fitted reliably with \FAT. We include the unreliable fits in this analysis for several reasons.  They serve as an illustration of why such galaxies should not be fitted and show that the regularity of a velocity field is no guarantee for a reliable fit. It is also interesting to see where the comparison between 3D and 2D fits starts to deviate as a confirmation of the reliability limits proposed by \cite{Bosma1978} and \cite{Begeman1987}.\\
\subsubsection{Central Coordinates}
\indent First we will look at the central coordinates (Figure \ref{LVHISOverview2} panels a and b). We find that \FAT\  finds centres comparable to \rc\ and \df. The grey circle in Figure \ref{LVHISOverview2}  (panel b) for the $\Delta$ RA and $\Delta$ DEC plot delineates the ellipse with $\sigma_{\rm \Delta RA}$ as the horizontal axis and $\sigma_{\rm \Delta DEC}$ as the vertical axis (see Table \ref{tab:ParametersLVHIS}).  These are on average half of the FWHM for the observations in this sample. Figure \ref{LVHISOverview2} illustrates that the centres fitted by the 3D branch agree better with the \df\ fits than with the \rc\ fits, even better than \df\ agrees with \rc.\\
\indent The single outlier in RA is  HIPASS J0317-66. It turns out that in this case it is difficult determine a single centre. The centre found in HIPASS (see table \ref{LVHISTable}) disagrees with the \df, \rc\ and \FAT\ fit ($\delta=-66\degr29\arcmin37.8\arcsec$, $\delta=-66\degr30\arcmin4.0\arcsec$, $\delta=-66\degr29\arcmin52.1\arcsec$, $\alpha=3^{\rm h}18^{\rm m}17.61^{\rm s}$, $\alpha=3^{\rm h}18^{\rm m}17.13^{\rm s}$, $\alpha=3^{\rm h}18^{\rm m}9.47^{\rm s}$, respectively). The optical centre agrees with the centre found in the \rc\ fit but  a visual inspection of the data cube favours the centre found by \FAT. As this is a large and complex galaxy (panel b in Figure \ref{mom1}), and a more detailed inspection of the data does not favour one centre over the other, we ignore this mismatch in our interpretation.  \\ 
\indent For the systemic velocity (panel a)  we see that although the \FAT\ fits for some galaxies differ by more than a channel width from the 2D results, the scatter is comparable to that between \rc\ and \df. We therefore conclude that \FAT\ successfully fits the centres of these galaxies.  From this we estimate that a typical error on the systemic velocity is about a channel in real observations, and that that on the spatial coordinates is about a fifth of the beam.\\
\subsubsection{Position Angle \& Inclination}
\begin{figure*} 
   \centering
   \includegraphics[width=19 cm, angle=0]{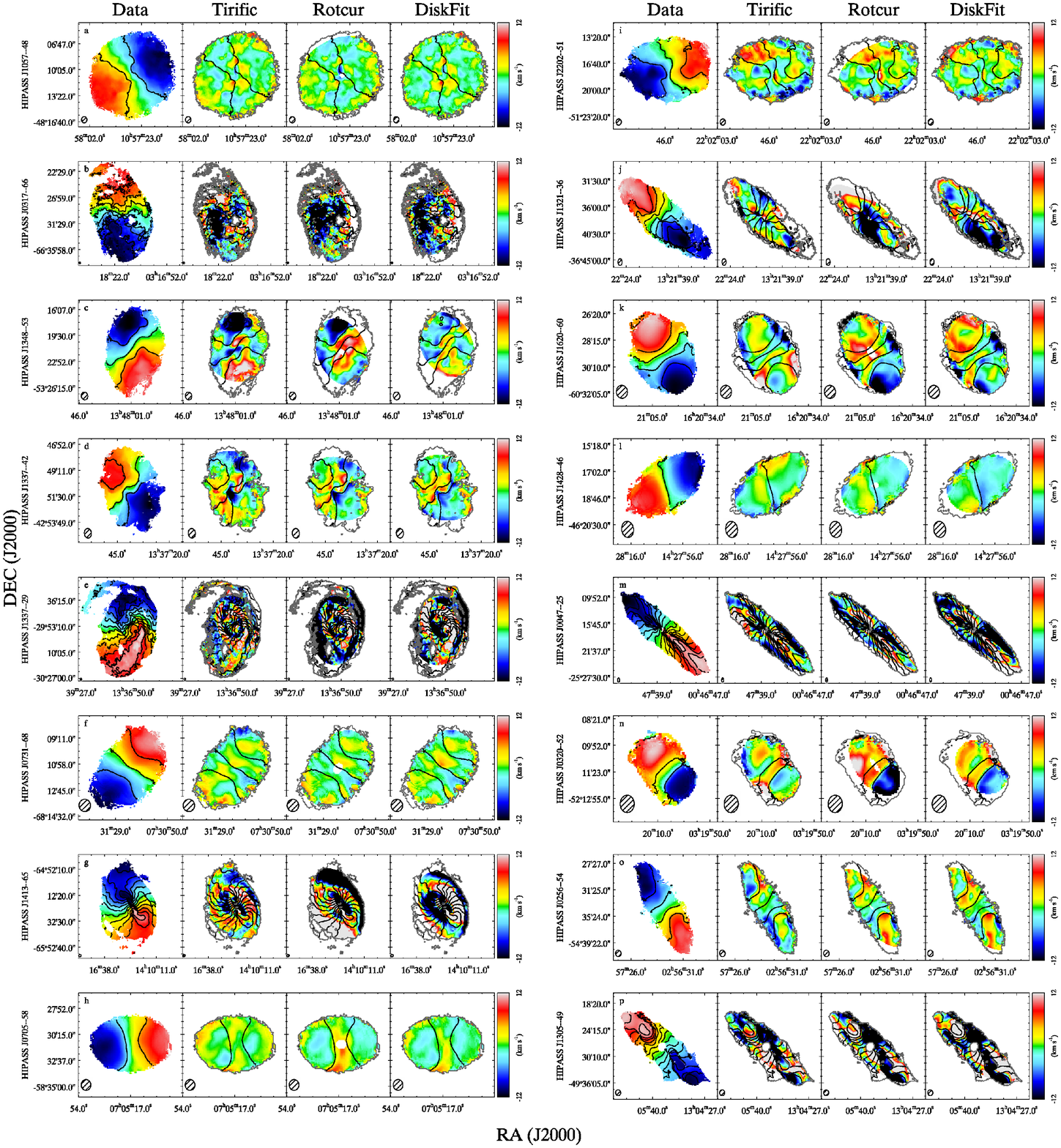} 
   \caption{Velocity field residuals of the subset of the LVHIS sample that has been fitted with \df, \rc\ and \FAT.  From left to right each row shows the velocity field of the Data, the residuals of \FAT, the residuals of \rc, the residuals of \df. The black contours always correspond to the contours of the data velocity field. They start at the minimum velocity in the velocity field and are drawn at intervals of  25 km s$^{-1}$. The grey solid line outlines the extent of the data velocity field.}
   \label{mom1res}
\end{figure*}

\begin{figure} 
   \centering
   \includegraphics[width=8 cm]{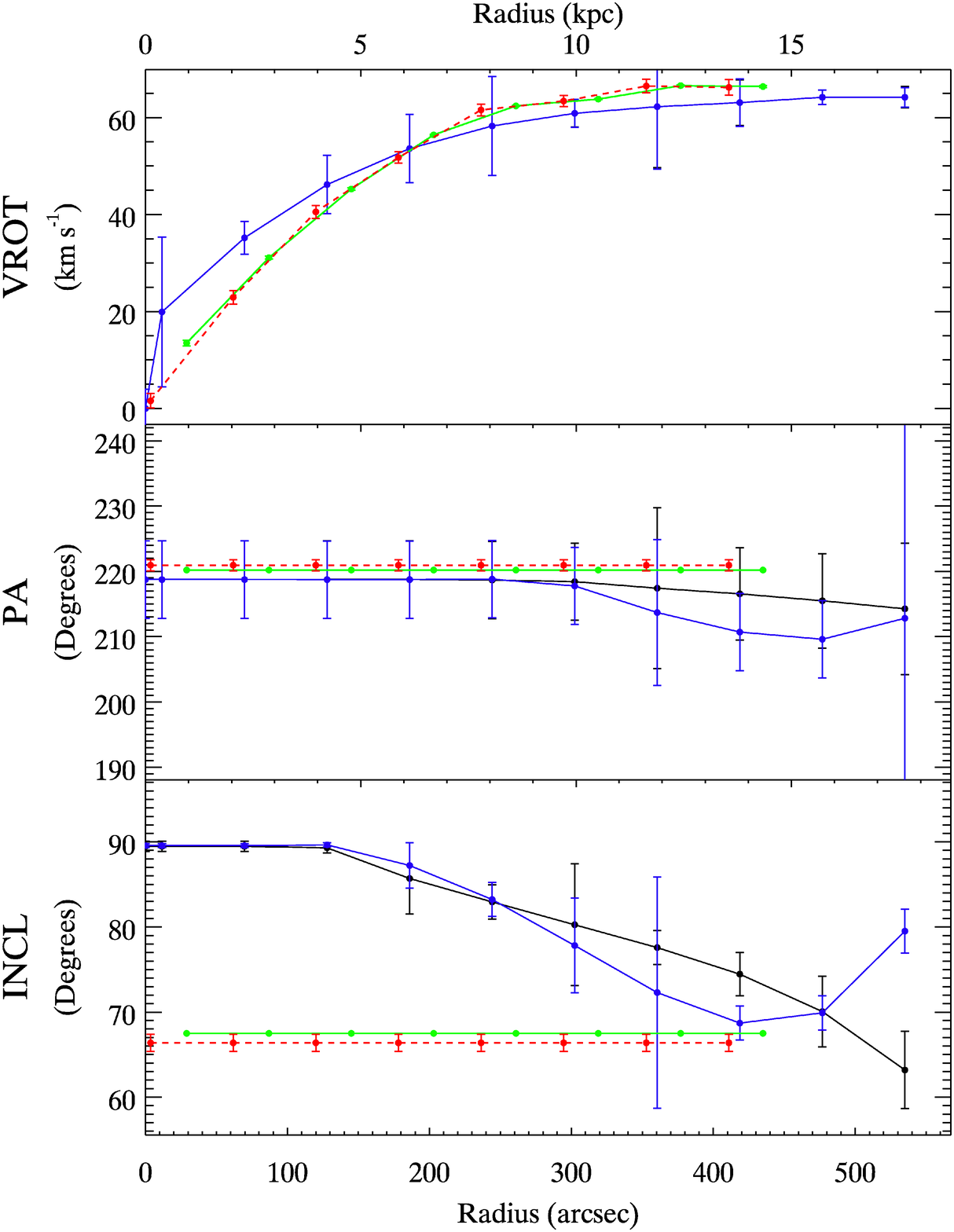} 
   \caption{HIPASS J0256-54: Parameters for the Rotation curve, PA and Inclination. Black + Blue: \FAT\ both sides independently, red: \rc\ green: \df. }
   \label{ESOinc}
\end{figure}

\indent  Similar to the central coordinates we see that \FAT\ provides a good fit for the PA compared to \rc\ and \df\ (Figure \ref{LVHISOverview2}, panel c). It is notable however that the scatter for this sample is  $\sim4^{\circ}$ and the mean absolute deviation  $\sim6.6^{\circ}$. This is due to a combination of errors and different fitting methods, as the position angle is allowed to change in the centre for the \rc\ fits but not for the \FAT\ or the \df\ fits.  The largest outlier in this plot is HIPASS J1413-65. For this galaxy the \rc\ fit to the PA does not vary whereas the \FAT\ model shows a variation in PA with several directional reversals, i.e. the PA increases, decreases and subsequently increases again. This leads to the large absolute mean deviation. A warp does seem to be favoured by the data (see Figure \ref{mom1}, panel g) and the 3D model traces the emission in the data cube well.\\
\indent It is clear from Figure \ref{LVHISOverview2} (panel d) that the biggest discrepancies can be found when comparing the inclinations. However, the deviations in the line-of-sight velocities, i.e v$_{\rm rot} \times {\rm sin(i)}$ (panel f in Figure \ref{LVHISOverview2} ),  are similar in most models and the scatter in their difference is half a channel  ($\sigma_{\rm obs}$= 0.5 channel). This makes it clear that \FAT, \df\ and \rc\ all solve for the degeneracy between rotational velocity and inclination in a different way. The outlier is a large edge-on galaxy (HIPASS J1305-49, Figure \ref{mom1} panel p) for which the 2D codes fit an inclination 10\degr\ lower than does \FAT. However, as this happens at high inclination the rotation curve is hardly affected leading to a large offset in the line-of-sight velocity. \\
\indent  From panel d in Figure \ref{LVHISOverview2} it can be seen that \df\ and \rc\ overall agree on inclination whereas \FAT\ clearly deviates  at all inclinations. Given the results in section \ref{AG}, we cannot exclude systematics errors in \FAT\ for galaxies smaller than 8 beams in diameter even in idealised tests. Hence, we will focus on the galaxies in the range reliably fitted by \FAT\ in what follows. \\
\indent Relative to 2D methods \FAT\ stretches the inclination range that is fitted. At low inclination \FAT\ finds inclinations lower than the 2D methods and at high inclination ones that are higher (see Figure \ref{mom1}). Note that only the velocity fields for the galaxies fitted with \df\ are shown in Figure \ref{mom1} as the other galaxies are more complex and the interpretation of the velocity field is not straight forward. In these velocity fields \FAT\ outer rings do not always correspond to the actual map because the fitting is done on an unmasked cube and can go below the noise in certain cases. At the low-end of the inclination range considered here, the ellipses and velocity fields show a lower inclination for the \FAT\ fits as they are more circular. It is hard to differentiate by eye between the different models and therefore we also present the residual velocity fields in Figure \ref{mom1res}.\\
\indent These residual fields emphasise that the codes solve for the  degeneracy between inclination and rotation in different ways. Even though the inclinations differ significantly, the residuals are similar. However, is the 3D method doing better? In order to address that question we will examine an individual galaxy in more detail. \\
\begin{figure*} 
   \centering
   \includegraphics[width=16 cm]{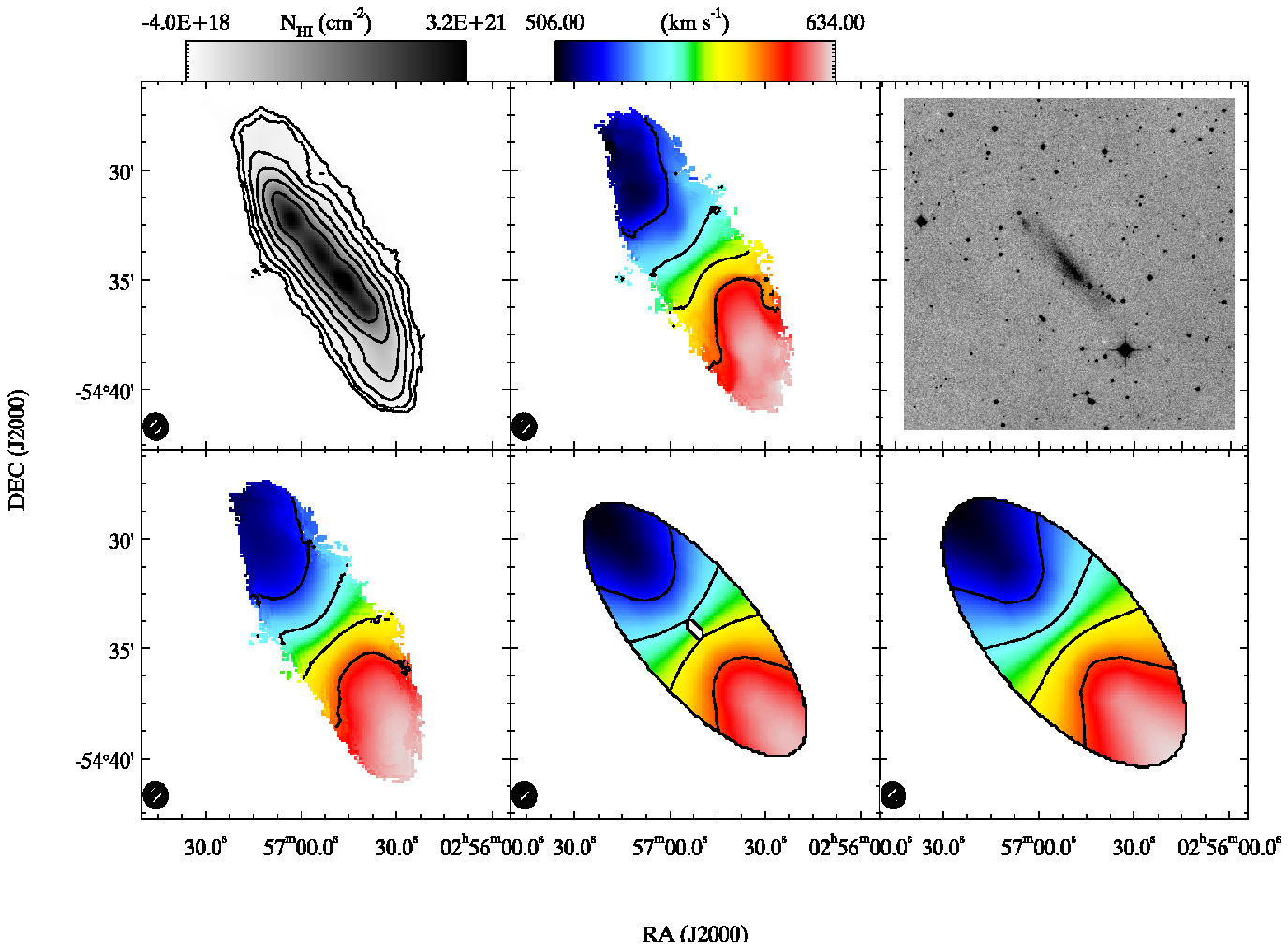} 
   \caption{HIPASS J0256-54: Top row from left to right: Integrated moment map, Gaussian-Hermite velocity field, Infrared image from DSS. Bottom row left to right: \FAT\ Gaussian-Hermite velocity field, \rc\ velocity field, \df\ velocity field. The black contours in the integrated moment map are at 6, 12, 24 .... etc. $\times10^{19}$ cm$^{-2}$ and for the velocity fields at 506, 531, 556, 581, 606 km s$^{-1}$. This galaxy is in also shown in panel o in Figures \ref{mom1} and \ref{mom1res}}
   \label{eso154-g23}
\end{figure*} 
\indent  Figure \ref{eso154-g23} shows the velocity field, an integrated moment map and an infrared image of HIPASS J0256-54. The difference in inclination between the  \df\ and \rc\  fits with \FAT\ is  17$^{\circ}$ and 16$^{\circ}$, respectively. A closer look at the inclinations (Figure \ref{ESOinc}) of the models shows that the \df\ and \rc\ fits are flat at $\sim$ 67$^{\circ}$ inclination whereas the 3D model is warped and varies from 89$^{\circ}$ in the centre to $\sim$70$^{\circ}$ in the outer parts. It is clear from the optical image and the integrated moment map in Figure \ref{eso154-g23} that this galaxy is close to edge-on in its inner parts, indeed from the optical axis ratios an inclination $\ge 88^{\circ}$ (see Table \ref{LVHISTable}) is expected. From the residual velocity fields it seems that the \FAT\ fit has somewhat lower residuals (see Figure \ref{mom1res}, panel o) but not significantly so.\\ 
\begin{figure*} 
   \centering
   \includegraphics[width=16 cm]{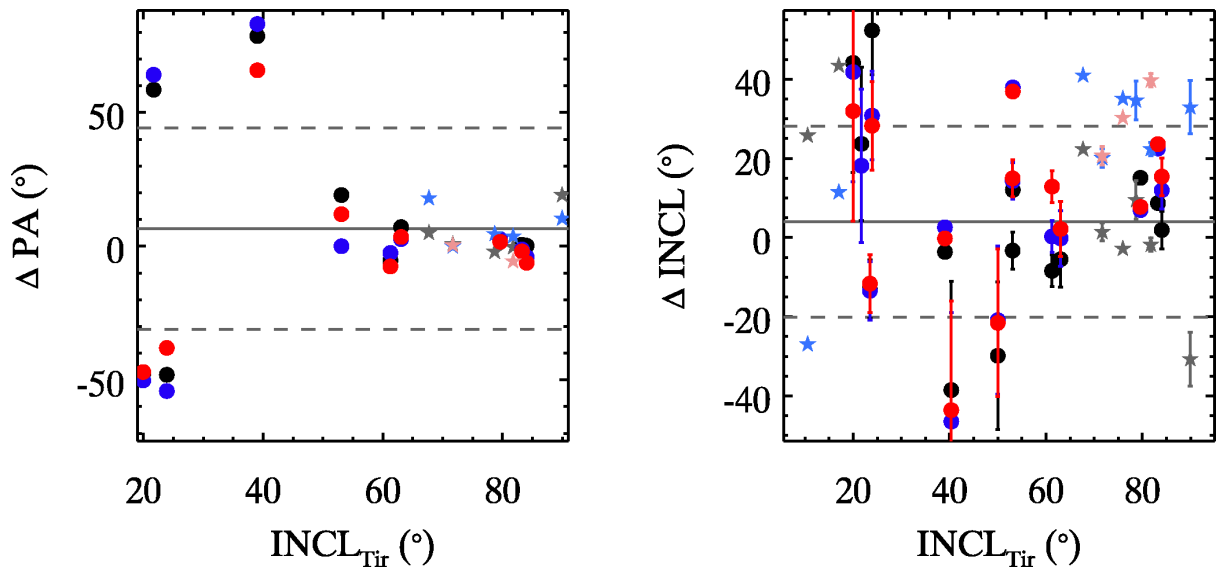} 
   \caption{Comparison between the PA and inclination of the optical values and the different fits, black - \FAT, blue - \rc, red - \df.  Stars are galaxies where the final \FAT\ model is in a part of parameter space that is considered unreliable.  The grey dash line denote the scatter on the deviations between literature and the \rc\ interactive fit. The optical values come from Table \ref{LVHISTable}.}
   \label{LVHISOverviewlit}
\end{figure*} 
\begin{table*}
   \centering
   \begin{tabular}{@{} lrrrrr @{}} 
	\hline
Fitting Code &  PA Mean Abs. Dev. & PA $\sigma$ & Inc.  Mean Abs. Dev & Inc. $\sigma$  & No. Gal. (Pa, Inc.) \\
\hline
\rc&4&40&8&24&10,14\\
\df&2&30&7&22&9,13\\
\FAT&7&38&4&24&10,14\\
   \hline
      \end{tabular}
   \caption{The mean absolute deviation and rms scatter of the difference between literature values and value found from the kinematic fitting for the different codes. The values of the model are extracted at the radius corresponding to the major axis radius found in the literature.}
      \label{tab:ParametersLVHISlit}
\end{table*}
\indent Figure \ref{LVHISOverviewlit} shows the difference between the literature values  for the inclination and PA, as stated in Table \ref{LVHISTable}, and all of the models. To ensure an appropriate comparison, the PA and inclination of the \HI\ observations were determined at the optical major axis radius found in the literature. \\
\indent From the left panel in Figure \ref{LVHISOverviewlit} it becomes immediately obvious that to determine a reliable PA  at low inclination, kinematical information is crucial as at the low inclinations there is a large difference between the optical PA and the PAs determined from the fits, regardless of the method. Below an inclination of  40\degr\ this difference is due to the inaccuracy of the optical PA estimates  \citep{Lauberts1982}. \\
\indent When we turn to the right hand panel of  Figure \ref{LVHISOverviewlit} it appears that at the high inclinations \FAT\ performs slightly better than the 2D methods.  However, this is based on only three galaxies (HIPASS J0047-25,  HIPASS J0256-54, HIPASS J1305-4) and the difference is not significant when compared to the overall scatter. An individual inspection of HIPASS J0047-25 and HIPASS J1305-4 (panels m and p in Figure \ref{mom1}) shows that the differences are created in a similar fashion as previously described for  HIPASS J0256-54. In the case of HIPASS J0047-25 the optical major axis is actually larger than the models and hence the warping in \FAT\ leads to a larger discrepancy compared to the 2D methods at the radius of the optical major axis. Overall however the fits are similar.\\
\indent Figure  \ref{LVHISOverviewlit}  shows that in the intermediate inclinations ($40^{\circ} < i < 70^{\circ}$) the 3D fits, on average,  agree with the optical values. This is also true for the 2D fits, however from Figure \ref{LVHISOverview2} it is clear that there is an average difference of $\sim$15\degr\  between the \FAT\ and 2D inclinations. Once more this is caused by \FAT\ fitting warped distributions where the 2D methods fit flat disks. One example of this is  HIPASS J0705-58 (Figure \ref{mom1}, panel h). From this velocity field it is clear that beam smearing, warps and a larger extent of the model can easily explain these differences. 
It appears the methods do equally well in this range when compared to the optical information available.\\
\indent At the low end of the low inclination range ($< 40^{\circ}$)  it is clear that the 3D method prefers lower inclinations compared to the 2D methods  (Figure \ref{LVHISOverview2}, panel d) and that in general both the 2D and 3D methods find inclinations lower than the literature values (Figure \ref{LVHISOverviewlit}).  Of the reliably fitted galaxies, four agree well between the different methods but one does not. This galaxy, HIPASS J1348-53, has a difference of more than $20^{\circ}$ between the 2D and 3D methods. It clearly  has an optically determined inclination that is much higher than suggested by the velocity field (76$^{\circ}$, Figure \ref{mom1}, panel c) making it impossible to differentiate between the methods.\\
\subsubsection{Rotation curves}
\begin{figure*} 
   \centering
   \includegraphics[width=17 cm]{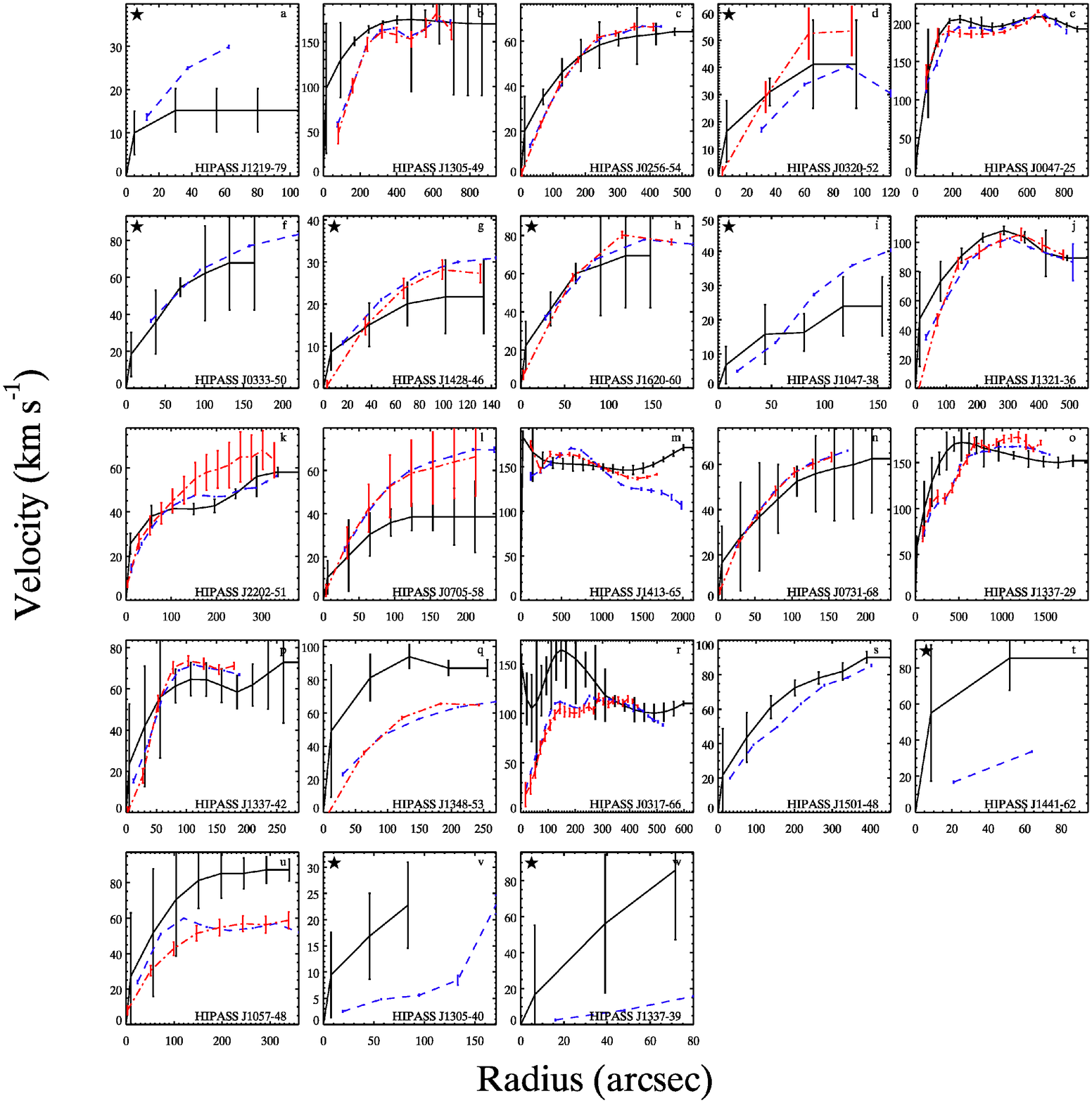} 
   \caption{Rotation curves for all the galaxies in the LVHIS Sample. The plots are ordered according to their \FAT\ inclination, with the lowest inclination in the bottom left. Black lines: \FAT, Red Lines:\df, Blue lines:\rc. The error bars are the formal errors derived by \df\ and \rc\ for the red and blue lines respectively. For \FAT\ the error is an empirical estimate derived from the regularisation process (see  $\S$ \ref{fittingProcedure}. Stars indicate curves outside the reliably fitted range.}
   \label{RCs}
\end{figure*} 
As tilted-ring modelling is pre-dominantly used for the extraction of rotation curves, we want to focus on how well the rotation curve is retrieved for each galaxy. Figure \ref{RCs} shows the rotation curves for the 23 galaxies in our sample fitted by \FAT. Here the black lines show the rotation curves derived with \FAT, the blue dashed lines derived with \rc\ and the red dot-dashed lines derived with \df. We see that in general, the \FAT\ rotation curve is consistent with those determined by the 2D methods albeit sometimes with an offset. Comparing these plots to panel d in Figure \ref{LVHISOverview2} it is evident that these offsets are a result of the different inclinations found by the 2D methods.\\
\indent In general the fitted rotation curves seem reasonable within their errors.  A possible difference between the 2D methods and \FAT\  is that in several cases the rotation curve found with the 3D method rises faster in the inner point. However, this also occurs in the inner most point of the rotation curves of the intermediate mass artificial galaxies (see Figure \ref{fakerotcurs}) and is therefore, at least partially, an artefact of the code. Again, \tiri\ interpolates linearly between the single ring radii. Any quantity with a high curvature is hence ill-fitted. This results in an overshooting at the fitted nodes.\\
\indent It is remarkable that also in this sample  large deviations start to occur  when there are less than eight beams across the major axis. This would imply that the methods for retrieving rotation curves from velocity fields have significantly improved since the work of \cite{Bosma1978}. However, a more detailed analysis than provided in this paper is required to confirm this.\\
\section{Discussion}\label{discussion}
\begin{figure*} 
   \centering
   \includegraphics[width=8 cm]{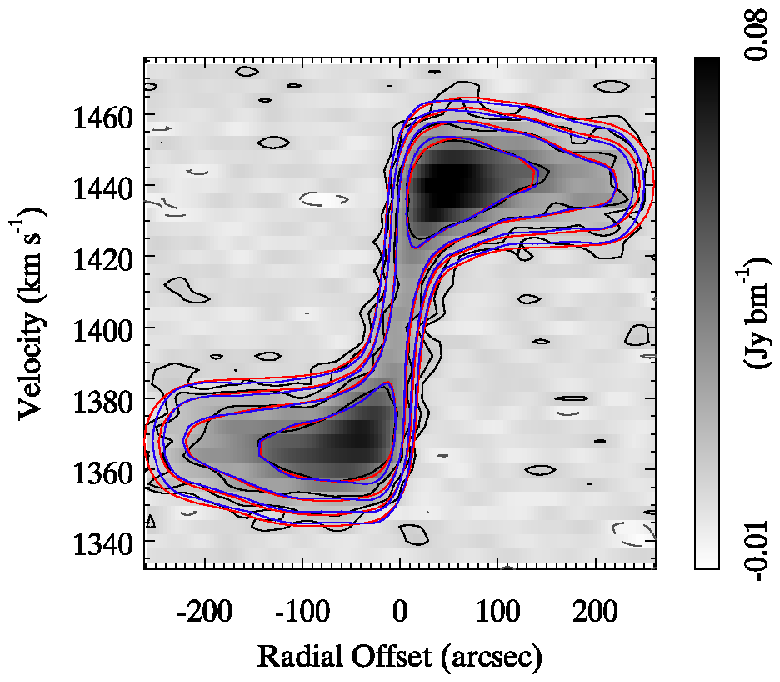} 
     \includegraphics[width=8 cm]{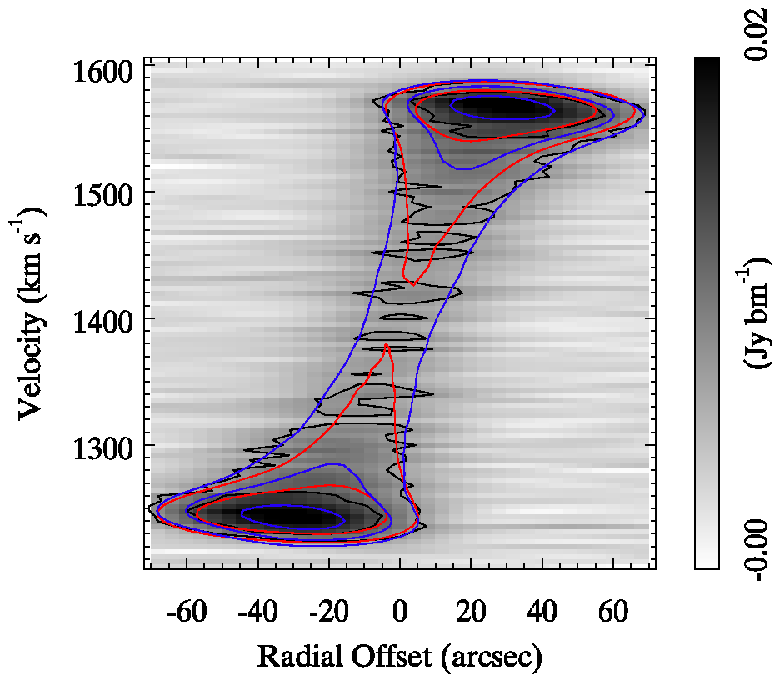} 
   \caption{The PV Diagrams for the artificial massive galaxy at 10 degrees inclination (left) and the massive galaxy with 4 beams across the major axis (right). Contours are at 2.5, 5, 10, 20 $\times$ 10$^{19}$ cm$^{-2}$. Blue contours are the input model, Black the input model plus noise and the red contours the final fit.}
   \label{PVfailed}
\end{figure*}

\indent We have shown that the code presented in this paper can, for the first time\footnote{While this paper was under review Bbarolo, another code capable of such fits, was released \citep{diTeodoro2015}.}, successfully fit tilted-ring models to \HI\ data cubes in a fully automated manner over a large range of projected and physical sizes. It works well for dwarf galaxies, large spirals and galaxies in between as long as they are resolved by at least eight beams across the major axis and in the inclination range 20\degr-90\degr. In order to achieve a successful fit all the code requires is the data cube of the galaxy. It will then go on to determine initial estimates \citep[through a run of SoFiA,][]{Serra2015}, the extent of the model, fit, adjust and smooth tilted-ring models without any input or adjustments from the user.\\
\indent The limits stated here pertain to fits without any priors on the kinematical parameters - in particular, it is likely that galaxies with even fewer  resolution elements across the major axis could be reliably fit if the inclination were fixed to the optical value \citep[e.g.][]{diTeodoro2015}.\\
\indent The aforementioned limits on the size of a galaxy's major axis relative to the beam and on its inclination in order to obtain a reliable FAT model are not hard, but will depend on other properties such as, for example, S/N and the shape of the rotation curve.  Additionally it must be emphasised that these are the limits for \FAT\ which is a fully automated code. This does not mean that 3D methods cannot fit outside these limits. However, in the case of low inclination it is very hard to see how one would separate incorrect fits from correct fits as the minimum in the $\chi^2$ landscape becomes increasingly more shallow. This is easily seen in Figure \ref{PVfailed} where the left panel shows the position velocity (PV) diagram for the massive artificial galaxy at 10 degrees inclination (black contours, greyscale), the same galaxy without noise (blue contours) and the fit (red contours) from \FAT. Even though the fitted rotation curve is, on average, almost 20 channels off, the modest deviation in inclination results in similar observed velocities.\\  
\indent On the other hand, for the amount of beams across the major axis there are differences. The right hand panel of Figure \ref{PVfailed} shows a PV-diagram along the major axis of the massive artificial galaxy with four beams across the major axis. The contours are the same as in the left panel. Even though the error in the rotation curve is much smaller for this galaxy, clearly the fit deviates from the input artificial galaxy without noise. However, this difference is hidden by the noise in this artificial galaxy. This implies that this lower limit for  number of resolution elements across the major axis can be reduced with improved fitting techniques or by increasing the signal to noise. Additionally the requirement of eight beams across the major axis is overly strict as the models fitted to a galaxy with seven  beams across the major axis end up with the same extent. This means that by merely improving the determination of the extent of the model the code can be improved to fitting galaxies with merely seven beams across the major axis. Furthermore, the choice of different interpolation methods to determine the inter-ring behaviour of parameters is also likely to further reduce the number of beams across the major axis required for a reliable fit.\\
\indent The requirement of eight beams across the major axis can probably also be pushed down by careful manual adjustment of the model, the introduction of priors (e.g. an optical inclination)  or a parametrisation of the galaxy model. As in the last case the degrees of freedom would be reduced this would make a more reliable fit but also the information that can be retrieved would be severely reduced, e.g. if one would use an analytical function for the shape of the rotation curve the maximum velocity could be retrieved and possibly the steepness of the rising part but investigations into rotation curve shapes would be ruled out. However, such methods and limits require further investigation.\\
\indent For the specific future of \FAT\ one could easily imagine replacing the first fit of a flat disc with a parameterised model in order to gain speed and accuracy. For galaxies with less than eight beams across the major axis this would also be the final fit and for better resolved galaxies \FAT\ would go on to fit an unconstrained model as it does in the current version. A concern here would be how much the initial parameterised model would steer the final results to this first model.\\
\begin{figure*} 
   \centering
   \includegraphics[width=16 cm]{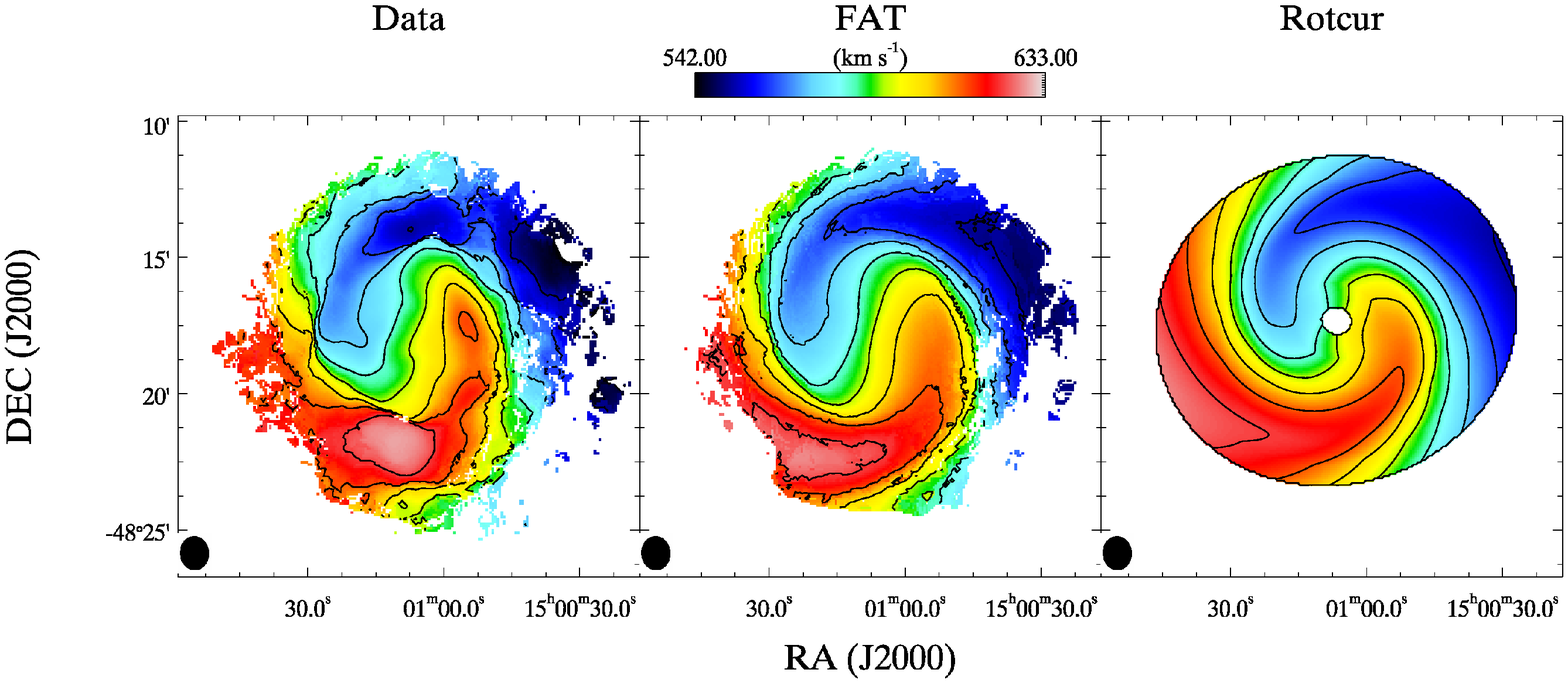} 
   \caption{The velocity fields for the galaxy HIPASS J1501-48. This galaxy is the most extremely warped galaxy in the sample. From left to right it shows the data, \FAT, \rc. Black contours appear at 10 km s$^{-1}$ intervals.}
   \label{VelsESO 223-G009}
\end{figure*} 
\indent The  \FAT\ can deal with extreme cases of warps. Figure \ref{VelsESO 223-G009} shows the velocity fields of the most extremely warped galaxy in the LVHIS sample, HIPASS J1501-48. This figure demonstrates the strength of tilted ring fitting in analysing  \HI\ observations with complex but regular structures. This galaxy also illustrates the problem of retrieving rotation curves in low inclination galaxies. Even though both the 2D and 3D methods provide good fits to the data the derived rotation curves (shown in Figure \ref{RCs}, panel s) are discrepant. This is due to the fact that in the \rc\ model the inclination is kept flat whereas \FAT\ finds a modest warp.\\ 
\indent \FAT\ fails to accurately reproduce an observed data cube in two cases, despite the fact that the code flags the fit as being successful. Currently the code does not check the models for physical consistency and this could be a way to identify cases that need to be inspected. \\
\indent When comparing the retrieved inclinations of the fits to optical inclinations it is unclear whether \FAT\ retrieves the inclination better than the 2D methods (Figure \ref{LVHISOverviewlit}) at high inclinations. In two cases it does, but in one case it does not. However, this latter case (HIPASS J0047-25) is a special one where the optical disc extends beyond the \HI\ layer. Other galaxies in our sample are either too small or complex and therefore this result is only tentative.\\
\indent In order to get a better estimate of when to use the 3D method and when to use the 2D method in the WALLABY kinematics pipeline it will be necessary to determine which of the two methods is more suitable for fitting observations of barely resolved galaxies, i.e. galaxies with less than eight beams across the major axis. The easiest way to investigate this  is to construct a database of artificial galaxies, larger than the one presented here, that can be fitted with the two branches of the pipeline. In this manner we know the input that we are trying to retrieve and we can accurately determine which method performs better. Such an investigation will also provide a much better idea of the errors on the final fits and whether there are other areas of the parameter space where one method performs much better than the other. Unfortunately, a simple grid with some variations in parameter space leads quickly to a set of thousands of galaxies. The current version of \FAT\ is not quick enough to deal with this in any reasonable time frame. A new version which can be run on high performance clusters is in development and will be tested on such a dataset.\\
\indent  From the tests on artificial galaxies we can get a lower limit of the errors in the \FAT\ fits. If we take the rms of the difference between the input and output we get a minimal error (column 1, Table \ref{tab:Errors}). As the artificial database lacks significant asymmetries and a large number of galaxies this scatter can be seen as the minimal error that the code will always provide. These are a gross underestimate of the real errors. A crude upper limit can be estimated from the scatter in the differences between the 2D  methods and \FAT\ (column 2, Table \ref{tab:Errors}). This would be a conservative estimate of the error as it includes galaxies in a range where 2D approaches are unsuitable.  In order to check this reasoning we compare these values to errors provided on tilted ring fits in the literature.\\
\begin{table*}
   \centering
   \begin{tabular}{@{} lccccc @{}} 
	\hline
Parameter &  Artificial Database & LVHIS Sample & \cite{deBlok2014} & \cite{Kamphuis2013} & \cite{Gentile2013}\\
 & & & (using \rc)& &\\
\hline
No. Galaxies & 46 & 14 & 1 & 1 & 1\\
RA (beam)& 0.02&   0.8 &  & 0.22  & \\
DEC (beam)& 0.01&   0.35& &   0.31& \\
V$_{\rm sys}$ (channel) & 0.09 &   0.9&  1.1&   1&  \\
PA ($^{\circ}$)&  0.3&   4.4&  0.6&   1&  2\\
Inclination ($^{\circ}$)&  0.8 &   6.3 & 1.5-2.5 &  1.3 & 2\\
V$_{\rm rot}$ (channel) &  0.5&   2.4& 2.5 & 1.2& 1.1 \\
   \hline
      \end{tabular}
   \caption{Errors on the tilted ring fit as estimated from the artificial database, the 2D to 3D comparison and in recent literature.}
      \label{tab:Errors}
\end{table*}
\indent Recently, several individual galaxies have been modelled with \tiri\ \citep{Gentile2013,Kamphuis2013,deBlok2014}. In these models the errors are estimated by varying each parameter individually until a comparison, by eye, between the data and the model clearly shows discrepancies. The errors from these publications are shown in the last three columns of Table \ref{tab:Errors}. These confirm that the scatter in the artificial galaxies is a lower limit on the error, whereas the scatter in the LVHIS sample agrees with these conservative errors. The exception to this is the large scatter in inclination  and PA that is found in the LVHIS sample. This is due to the fact that the major differences between the codes stem from how they deal with PA and inclination. \df\ always fits flat inner discs, whereas \FAT\ and \rc\ use various degrees of warping. Additionally, \df\ and \rc\ fit the whole galaxy at once whereas \FAT\ divides the galaxies in two along the minor axis. This does not happen for the rotation curve or the central coordinates because similar symmetry requirements are applied to these parameters. Therefore, the scatter on the PA and inclination is likely to be an overestimation of the typical error on these parameters. We conclude, that when similar fitting methods are applied, the scatter between the 2D methods and \FAT\ gives a good estimate of the typical error on a \FAT\ fit. Future investigations will aim at an independent determination of the errors, via bootstrapping or Markov Chain Monte Carlo (MCMC) methods.
\section{Summary and Conclusions}\label{summary}
\indent In this paper we present a fully automated code for fitting tilted-ring models (the Fully Automated Tilted Ring Fitting Code or \FAT) to \HI\ data cubes.
This code is an IDL wrapper around \tiri\  \citep{Jozsa2007} that uses \sofia\ \citep{Serra2015} to obtain initial parameter estimates. It is intended for use as the 3D branch of  a secondary data product pipeline for the ASKAP all sky survey WALLABY \citep{Koribalski2012} and its northern counter part WNSHS. Besides the 3D code itself we also present the concept for this pipeline of which the 2D branch will be presented in an accompanying paper \citep{Oh2015}.\\
\indent In order to test the code we have created a set of 52 artificial galaxies and selected 25 galaxies from the nearby galaxy survey LVHIS \citep{Koribalski2010}. Test runs on artificial galaxies indicate an inclination between 20\degr and 90\degr and an extent of eight beams across the major axis as a highly conservative lower limit to produce an acceptable fit. Further investigations will likely push down these limits by a large factor. Even now, galaxies at the lowest inclinations are rejected only because the large errors in the rotation velocity render the fits unusable for rotation curve analyses (but not, e.g. for studies of the surface brightness distribution).\\
\indent Out of a sample of 25 LVHIS galaxies 23 are fitted successfully in the sense that their models trace the observed emission when comparing data cubes to model cubes. The failed fits are due to residual artefacts in the observations and complex emission patterns of the \HI. However, the current version of the code fails to flag the unsuccessful fits and they need to be identified manually. As the models for these galaxies differ significantly from the data this could still be achieved without the need for any additional observations. \\
\indent From the fitting of 23 galaxies in the LVHIS sample we show that 2D fitting techniques underestimate the inclinations of galaxies at high inclination. As in our current sample there are only three galaxies at high inclination that are fitted with the 2D methods as well as \FAT\ it is unclear whether this is a systematic effect. Additionally, as this happens at high inclinations where the change of sin($i$) is small as a function of $i$ this does not significantly affect the amplitude of the derived rotation curve.\\
\indent \FAT\ retrieves several parameters related to the brightness distribution such as the scale height and brightness profile. When the scale height is significantly smaller than the beam this results in an overestimation of the scale height, and we conclude that hence the scale height cannot be estimated using \FAT\ for our test sample. The current tests are not sufficient to identify what happens when the scale height is resolved.\\
\indent Errors on the individual parameters remain ill understood.  However, a comparison with recent modelling work shows that the scatter in the sample of LVHIS galaxies is a good indication of realistic errors. In order to investigate the open questions on the code's performance, a run on a large artificial database of galaxies, where the exact structure of the galaxies is known, is required. Unfortunately, the current code is not fast enough to deal with the requisite number of fits. Hence, the next step is to adapt the code to run on high performance computing facilities. \\
\section*{Acknowledgements} 
We would like to thank an anonymous referee, T. Westmeier, S. Haan, W.G.J. de Blok and N. Giese for their useful comments on earlier drafts of this paper. PK would like to thank the Humboldt foundation for their support during his stay in Germany. KS acknowledges support from the Natural Sciences and Engineering Council of Canada. The Australia Telescope Compact Array is part of the Australia Telescope National Facility which is funded by the Commonwealth of Australia for operation as a National Facility managed by CSIRO. This research has made use of the NASA/IPAC Extragalactic Database (NED) which is operated by the Jet Propulsion Laboratory, California Institute of Technology, under contract with the National Aeronautics and Space Administration.\\

\bibliographystyle{mn2e}   
\bibliography{references}
\label{lastpage}
\end{document}